\documentclass[structabstract]{aa}  
\usepackage{graphicx}
\usepackage{txfonts}
\begin{document}
\newcommand{\escs}{erg\,s$^{-1}$cm$^{-2}$sr$^{-1}$}
\newcommand{\cmt}{cm$^{-3}$}
\newcommand{\cmd}{cm$^{-2}$}
\newcommand{\h}{H$_2$}
\newcommand{\kms}{\,km\,s$^{-1}$}
\newcommand{\um}{$\,\mu$m}
\newcommand{\lam}{$\lambda$}
\newcommand{\wat}{H$_2$O}

   \title{Mapping water in protostellar outflows with \textit{Herschel}\thanks{Herschel is an ESA space observatory with science instruments provided by European-led Principal Investigator consortia and with important partecipation from NASA}}

   \subtitle{PACS and HIFI observations of L1448-C}

   \author{B. Nisini\inst{1} 
   \and G. Santangelo\inst{1}
   \and S. Antoniucci\inst{1}
   \and M. Benedettini\inst{2} 
   \and C. Codella\inst{3} 
   \and T. Giannini\inst{1}
   \and A. Lorenzani\inst{3}
   \and R. Liseau\inst{4} 
   \and M. Tafalla\inst{5} 
   \and P. Bjerkeli\inst{4} 
   \and S. Cabrit\inst{6}
   \and P. Caselli\inst{7,3} 
   \and L. Kristensen\inst{8} 
   \and D. Neufeld\inst{9} 
      \and G. Melnick\inst{10}
   \and E. F. van Dishoeck\inst{8,11} 
}

   \institute{
INAF - Osservatorio Astronomico di Roma, Via di Frascati 33, 00040 Monte Porzio Catone, Italy:
              \email{nisini@oa-roma.inaf.it} 
\and 
INAF - Istituto di Astrofisica e Planetologia Spaziali, via Fosso del Cavaliere 100, 00133, Roma, Italy
\and 
INAF – Osservatorio Astrofisico di Arcetri, Largo E. Fermi 5, 50125 Firenze, Italy
\and 
Department of Earth and Space Sciences, Chalmers University of Technology, Onsala Space Observatory, 439 92 Onsala, Sweden
\and 
IGN Observatorio Astron\'{o}mico Nacional, Apartado 1143, 28800 Alcal\'{a} de Henares, Spain
\and 
LERMA, Observatoire de Paris, UMR 8112 of the CNRS, 61 Av. de L'Observatoire, 75014 Paris, France
\and
School of Physics and Astronomy, University of Leeds, Leeds LS2 9JT
\and
Leiden Observatory, Leiden University, PO Box 9513, 2300 RA Leiden, The Netherlands
\and 
Department of Physics and Astronomy, Johns Hopkins University, 3400 North Charles Street, Baltimore, MD 21218, USA
\and 
Harvard-Smithsonian Center for Astrophysics, 60 Garden Street, MS 42, Cambridge, MA 02138, USA
\and
Max Planck Institut for Extraterestrische Physik, Garching, Germany}

   \date{Received September 15, 1996; accepted March 16, 1997}

 
  \abstract
   {
   Water is a key probe of shocks and outflows from young stars, being extremely sensitive to both 
   the physical conditions associated with the interaction of supersonic outflows with the ambient medium,
   and the chemical processes at play.
}
   {Our goal is to investigate the spatial and velocity distribution of \wat\ along outflows, 
   its relationship with other tracers, and its abundance variations. In particular, this
   study focuses on the outflow driven by the low mass protostar L1448-C, 
   which previous observations 
   have shown
    to be one of the brightest \wat\, emitters among 
   the class 0 outflows.}
{To this end, 
	maps of the o-\wat\, 1$_{10}$-1$_{01}$
    and 2$_{12}$-1$_{01}$ transitions taken with the \emph{Herschel}-HIFI and PACS instruments, respectively, are presented. 
    For comparison, complementary maps of the CO(3-2) and SiO(8-7) transitions, obtained at the JCMT, 
     and the H$_2$ S(0) and S(1) transitions, taken from the literature, have been 
also used. Physical conditions 
     and \wat\, column densities have been inferred
with the use of LVG radiative transfer calculations.}
  {The water distribution appears clumpy, with individual peaks corresponding to shock spots 
   along the outflow. The bulk of the 557 GHz line is confined
   to radial velocities in the range
   $\pm$10-50 \kms, but extended emission at extreme velocities (up to v$_r \sim$ 80 \kms) is
   detected and is associated with the L1448-C extreme high velocity (EHV) jet. The \wat\,1$_{10}$-1$_{01}$/CO(3-2)  ratio shows strong variations as a function of velocity
     that likely reflect different and changing physical conditions in the gas responsible for the 
   emissions from the two species. In the EHV jet, a low \wat/SiO 
   abundance ratio is inferred, that could indicate molecular formation from dust free gas 
   directly ejected from the proto-stellar wind. 
   The ratio between the two observed \wat\, lines, and the comparison with \h, indicate 
   \textit{averaged} $T_{kin}$ and $n(H_2)$ values of $\sim$ 300-500 K and 5\,$10^6$ \cmt\,  
   respectively, while a water abundance with respect
   to \h\, of the order of 0.5-1\,10$^{-6}$ along the outflow is estimated, in agreement 
   with results found by previous studies. 
   The fairly constant conditions found all along the
   outflow implies that evolutionary effects on the timescales of outflow propagation do not
   play a major role in the \wat\, chemistry. 
   }
    {The results of our analysis show that the bulk of the observed \wat\, lines comes from 
   post-shocked regions where the gas, after being heated 
   to high temperatures, 
    has been already cooled down to  
a few hundred K. 
   The relatively low derived abundances, however, call for some mechanism to diminish the \wat\, gas 
   in the post-shock region. Among the possible scenarios, we 
 favor \wat\, photodissociation, which requires the superposition of a low velocity non-dissociative shock with a fast dissociative shock able to produce a FUV field of sufficient strength.
   }
 

   \maketitle
%

\section{Introduction}

The earliest stages of star formation are characterized by strong mass loss, which 
is at the origin of observationally prominent phenomena, such as 
shocks and molecular outflows. 
The high velocity of the shocked gas, 
and the elevated gas temperature, strongly modify
the chemical composition of the gas. 
Depending upon the initial conditions, processes that modify the gas composition
include gas dissociation and ionization, high temperature chemical reactions
and dust grain reprocessing (e.g. Flower et al. 2010).
These processes produce observable signatures in the form of emission  
from 
specific molecular and/or
atomic lines, 
the study of which is
crucial, not only as a probe of the shock chemistry,
but also for understanding the complex interaction 
between wind/jet-shocks and large scale outflows.

Among the different tracers, lines of \h\, and CO are routinely used to
infer the physical conditions and the dynamics of shocked gas, while 
less abundant molecules, like 
SiO or CH$_{3}$OH, are sensitive to 
the chemical processes triggered in the shocked gas. 
In this framework, water can be considered a key molecule: in fact, the \wat\, relative
line intensities and its column density
are subject to large variations that are highly 
dependent on both the actual physical conditions of the gas but also on its
thermal and chemical history. This is because the water abundance strongly depends on both the 
mechanism of evaporation/freeze-out in grain mantles and the endothermic 
gas-phase chemical reactions that drive all free oxygen into water,  
as well as on
the relative timescales of these processes (e.g. Bergin et al. 1998; Flower 
\& Pineau des For{\^e}ts 2010).

Observations obtained with the  Infrared Space Observatory (\textit{ISO}) have been
the first to detect \wat\, 
emissions from states of relatively high excitation
($T_{kin} \sim$ 500-1500 K, e.g. Liseau et al. 1996; Ceccarelli et al. 1998; Nisini et al. 2000).
More recently, the SWAS and Odin satellites observed the fundamental o-\wat\, transition at
557 GHz in a sample of outflows (Franklin et al. 2008; Bjerkeli et al. 2009; Benedettini et al. 2002).
These observations probed cooler gas  
than had been observed with ISO,
but 
were able to resolve the line profiles for the first time, demonstrating the association
of water emission with the high velocity gas.
These studies provided the first determinations of the water abundance, 
yielding values in the range $\sim$ 10$^{-7}$ to $\sim$10$^{-4}$
and suggesting that the \wat\, abundance depends on both the
gas temperature and speed (Giannini et al., 2001; Franklin et al. 2008).
However, the strength of this conclusion was limited by the large beam sizes used in
these previous observations,  together with their limited spectral resolution and/or excitation coverage; 
these limitations made it difficult to associate enhanced abundances or broadened line profiles  
with specific regions along the outflows
or to infer whether these globally-averaged properties are really representative
of the physical and chemical conditions in specific regions of shock activity.

\textit{Herschel} (Pilbratt et al. 2010) represents the natural evolution for the study of \wat\, in protostellar 
sources, thanks to the combination of much improved spectral/spatial resolution
and sensitivity provided by the PACS (Photodetecting Array Camera and Spectrometer, Poglitsch et al. 2010) 
and HIFI (Heterodyne Instrument for the Far Infrared, de Graauw et al. 2010) instruments. 
In the framework of the ''Water In Star-forming regions with \textit{Herschel}" (WISH, van Dishoeck 
et al. 2011) key program, we have undertaken systematic PACS and HIFI observations
of young outflows in nearby clouds. 
Within this program, studies of individual shocks
have been published in Bjerkeli et al. (2011), Santangelo et al. (2012), 
Vasta et al. (2012) and Tafalla et al. (2012), while water maps of the L1157 and VLA1623 outflows
have been presented in Nisini et al. (2010) and Bjerkeli et al. (2012). 
All these studies complement observations at the central source position,  
which probe outflowing gas shocked in the inner jet and envelope cavity walls 
(Kristensen et al. 2012, Herczeg et al. 2011, Kaska et al. 2012, Goicoechea et al. 2012).


This paper will focus on  
PACS and HIFI mapping observations of the 
outflow from the class 0 source L1448-C (also named L1448-mm). This is a 
low luminosity (\textit{L} = 7.5 L$_\odot$;
Tobin et al. 2007) protostellar source located in the Perseus
Molecular Cloud (\textit{D}= 232 pc; Hirota et al. 2011), 
which drives a powerful and highly collimated flow 
that has been detected through interferometric CO and SiO observations (Guilloteau et al.
1992; Bachiller et al. 1995; Hirano et al. 2010). 
To the North, the L1448-C outflow interacts with two more compact flows 
originating from a small cluster of three young sources (L1448-NA, NB and NW, Looney et al. 2000).

Regions of shocked gas are seen along the entire outflow by 
means of near- and mid-IR
of molecular hydrogen emission (Davis \& Smith 2006; Neufeld et al. 2009; Giannini et al. 2011), 
which indicate 
the presence of a gas at a large range of temperatures, from $\sim$ 300 to more than 2000 K.
\textit{ISO} detected 
a far-IR spectrum rich  
in \wat\, and CO transitions
towards the L1448-C outflow
 (Nisini et al. 1999, 2000). The analysis of these lines
constrained their emission as coming from 
warm gas with an enhanced water abundance, as predicted 
by models for non-dissociative shocks.
\textit{SWAS} and \textit{Odin} detected the 557 GHz line but at a 
single-to-noise ratio that was
too low to characterize its emission 
kinematically. These studies, however, suggest that
this line might probe a colder water gas component whose abundance is less 
enhanced with respect to the warm gas. HDO emission 
at 80.6 GHz has also been detected
towards L1448-C, and is
associated with both the protostar and the shocked walls of the 
outflow cavity (Codella et al. 2010b).

Within the WISH program, the L1448-C outflow has been the subject of a detailed
study that includes, in addition to the mapping observations presented here,  
a survey of several lines at specific positions. 
In particular, \textit{Herschel}-HIFI observations of the central L1448-C source have been reported by
Kristensen et al. (2011), who detected prominent emission originating  from both
a broad velocity component, probably associated with the interaction of the
outflow with the protostellar envelope, and from the Extreme
High Velocity gas (EHV, the so-called ''bullets'') associated  
with the collimated molecular jet. Santangelo et al. (2012) discussed 
observations carried out
towards two specific shock spots, 
and showed that \wat\, line profiles change significantly
with excitation, indicating the presence of gas components
having different physical conditions. 

The main aims of this work will be to 
define the global morphological and kinematical 
properties of the \wat\, emission, 
in comparison with other standard outflow and shock tracers,
and to study abundance variations in the different shocked regions.
To this end, complementary CO(3-2) and SiO(8-7) maps of the same region
covered by the \textit{Herschel} observations will be presented and discussed.


\section{Observations}

\subsection{PACS observations}

Observations with the PACS instrument were performed on 27 February 2010 
(with observing identification number OBSID=1342191349). 
The PACS Integral Field Unit (IFU) in line spectroscopy mode
was used in chopping/nodding mode
to obtain a spectral map of the L1448 outflow centered 
on the \wat\ 2$_{12}$-1$_{01}$ line at 179.527\um\, (i.e. 1669.905 GHz, hereafter 
referred to as the ``179\um\, line''). 
The IFU consists of a 5$\times$5 pixel array
providing a spatial sampling of 9$\farcs$4/pixel, for a total field of view of 
47$\arcsec \times 47\arcsec$. The diffraction-limited FWHM beam size at 179\um\, is 12$\farcs$6.
The L1448 outflow region (about $5\arcmin \times 2\arcmin$
centered on the L1448-C(N) source, $\alpha$(J2000) = 03$^h$25$^m$38.4$^s$,
$\delta$(J2000) = +30$^o$44$\arcmin$06$\arcsec$) was covered through a single Nyquist
sampled raster map, arranged along the outflow axis. 
The \textit{Herschel} pointing accuracy is  
$\sim$ 2$\arcsec$.

The spectral resolution at 179\um\ is $R\sim$1500 (i.e. $\sim$210 km\,s$^{-1}$). 
The observation was performed with a single scan cycle, providing an integration 
time per spectral resolution element of 30 sec. The 
total on-source time for the entire map was 5670 sec.

The data were reduced with HIPE\footnote{HIPE is a joint development 
by the \textit{Herschel} Science Ground Segment Consortium, consisting of ESA, 
the NASA \textit{Herschel} Science Center, and the HIFI, PACS and
SPIRE consortia} v6.0, where they were flat-fielded and
flux-calibrated by comparison with observations of Neptune. The calibration uncertainty 
amounts to around 20-30$\%$, based on cross-calibrations with HIFI and ISO, and on continuum 
photometry (internal WISH report). Finally, in-house IDL routines were
used to locally fit and remove the continuum emission, and to construct an integrated 
line map. 

\subsection{HIFI observations}

\begin{figure}
\centering
\includegraphics[angle=-90,width=9cm]{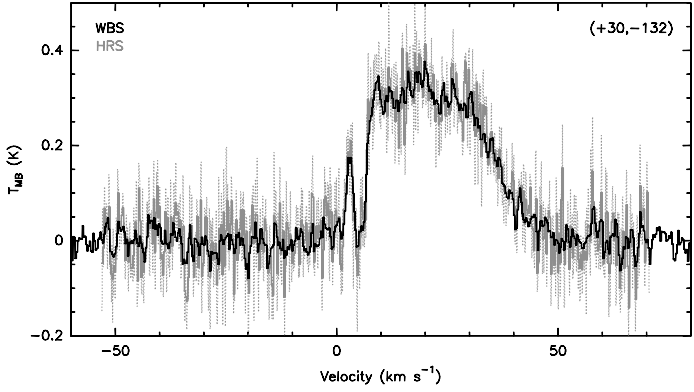}
\caption{Comparison between a WBS and HRS spectrum at a representative position 
in the outflow (offsets with respect to L1448-C are indicated). Full black lines 
show the WBS data, while the gray dotted and full lines 
show the HRS data 
before and after smoothing to
the WBS resolution, respectively.}
\label{hrs}%
\end{figure}

A region of $5\arcmin \times 2\arcmin$ oriented along the direction of the L1448 outflow 
(PA 164$\degr$) was mapped in the \wat\ 1$_{10}$-1$_{01}$ line at 556.936 GHz (i.e. 538.29\um, hereafter ``the 557 GHz line") with the HIFI instrument (de Graauw et al. 2010) on 19 August 2010 (OBSID: 1342203200). 
The On-The-Fly (OTF) mode
was adopted, with a distance between adjacent scans of 16$\arcsec$, 
slightly less than half the diffraction HPBW 
(which is 38$\arcsec$ at the observed line frequency).
The observations were performed in Band 1b 
with both the Wide Band (resolution 1.1 MHz) and High Resolution (resolution 0.25 MHz) Spectrograph backends (WBS and HRS, respectively), for a total on-source integration time of 3981sec. 
An inspection of the two sets of data showed that the HRS spectra
fail to provide additional information
on the line velocity structure, 
and, furthermore, result in an higher rms noise 
when smoothed to the resolution of the WBS data (see an example in Fig. \ref{hrs}).
Hence in this paper, only the WBS data have been used.
The data were reduced using HIPE v7, while further analysis was performed using the GILDAS\footnote{http://www.iram.fr/IRAMFR/GILDAS/}
package. Calibration of the raw data 
onto the $T_A$ scale was performed
by the in-orbit system, while the spectra were
converted to a $T_{mb}$ scale adopting a main beam efficiency $\eta_{mb}$=0.75
(Roelfsema et al. 2012).

Additional analysis consisted of baseline removal in each individual spectrum,
averaging of spectra taken during different cycles, and construction of 
a final data-cube sampled at a regular grid having a half-beam spacing.
Observations from the H and V polarizations were separately reduced: 
spectra from the two polarizations were acquired at slightly different 
coordinates (offset of $\sim$ 7$\arcsec$) that have been taken into account
in constructing the final regridded map.
The rms noise achieved in the final data-cube is typically of the order of $T_{mb} \sim$ 0.02 K 
in a 1 \kms\, bin. 

\subsection{JCMT-HARP observations}
Complementary  CO(3-2) and SiO(8-7) OTF maps were
obtained in January 2009 with the HARP-B heterodyne array (Smith et al. 2008) 
and ACSIS correlator (Dent et al. 2000) on
the James Clerk Maxwell Telescope (JCMT).
The rest frequencies are 345796.0 and 347330.6 GHz for CO(3-2) and SiO(8-7), respectively
(Pickett et al. 1998).
The mapped area was covered by consecutive scans 
in basket-weave mode at a position angle of 160$\degr$.
Each scan was offset by 29$\farcs$1 in the orthogonal direction,
and the signal was integrated every 7$\farcs$3 (half HPBW, about 14$\arcsec$) along the 
scan direction. 
We observed in standard position-switched observing mode, with an off-source 
position at (+140$\arcsec$, 0$\arcsec$), chosen to be 
devoid of sources and the presence
of high velocity gas.
Single maps were co-added and initial data cubes converted into
GILDAS format for baseline
subtraction and subsequent data analysis.
The resulting map is centered on $\alpha_{\rm J2000}$ =
03$^{\rm h}$ 25$^{\rm m}$ 38$\fs$9 $\delta_{\rm J2000}$ = +30$\degr$
44$\arcmin$ 05$\farcs$0, and it has dimensions of 300$\arcsec$ $\times$ 116$\arcsec$.

The observed bandwidth, 1 GHz, 
was sampled with 2048 channels for a 
spectral resolution of 488 kHz, which corresponds to 
0.42 km s$^{-1}$ at the observed frequencies. The spectra 
were smoothed to 1 km s$^{-1}$ 
resolution, to increase the sensitivity, and converted to  
the main-beam brightness temperature ($T_{\rm mb}$) scale adopting a 
main-beam efficiency ($\eta_{\rm mb}$) of 0.6.
The mean rms noise in $T_{\rm mb}$ is around 100 mK and 80 mK for CO(3-2) and SiO(8-7),
respectively.


\begin{figure*}
\centering
\includegraphics[angle=-90,width=12cm]{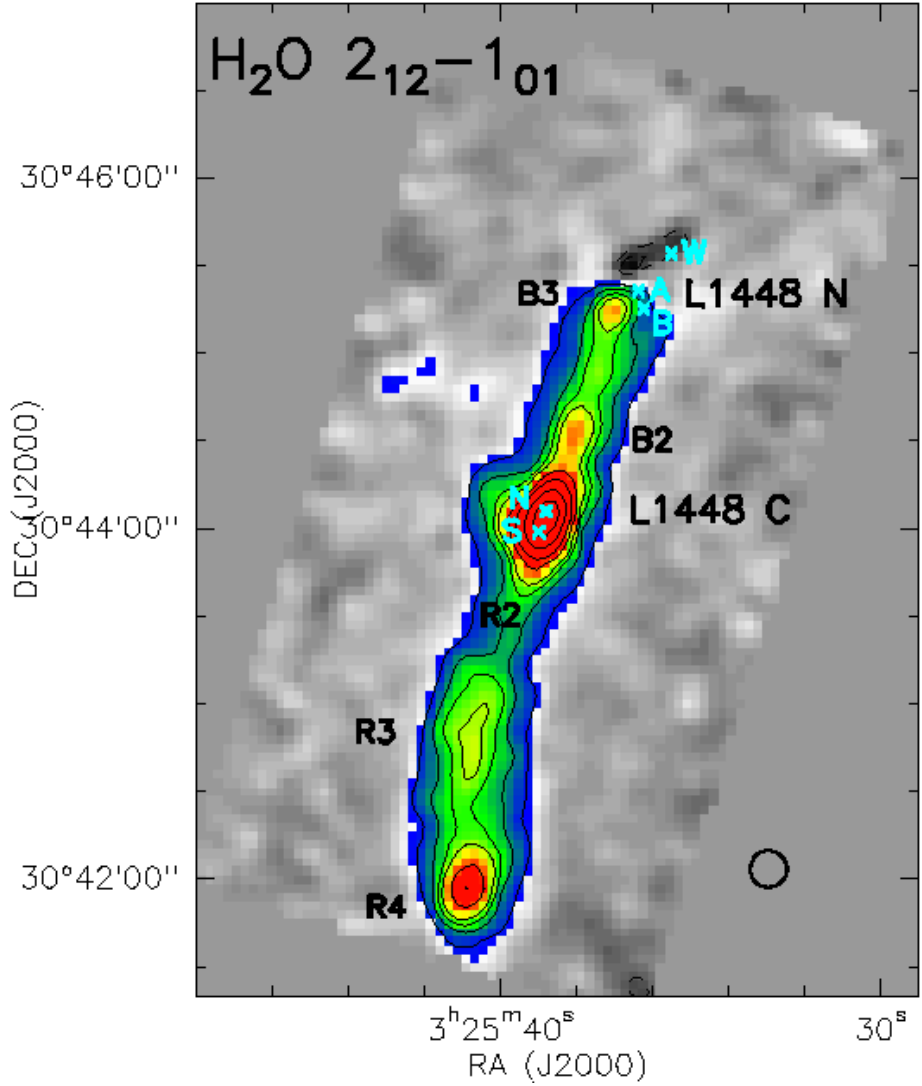}
\caption{Continuum-subtracted PACS map of the integrated H$_2$O 2$_{12}$-1$_{01}$ emission along the L1448 outflow. Sources in the region are indicated with crosses: L1448-C(S) and C(N) 
in the central region, and L1448 NA, NB and NW in the northern region. The diffraction-limited beam of 
FWHM 12.6$\arcsec$ is also indicated.
 The different emission peaks are labelled following the nomenclature adopted 
 by Bachiller et al. (1990) for individual CO peaks. The average rms noise in the
 observed region is of the order of 2$\times$10$^{-6}$ \escs\,. Contours
 are drawn at 3$\sigma$, 6$\sigma$,9$\sigma$, 12$\sigma$, 16$\sigma$, 30$\sigma$,
 50$\sigma$ and 75$\sigma$. Negative contours are indicated by
 dashed lines.}

\label{pacs}%
\end{figure*}

\section{Results}
\subsection{\wat\, morphology}

The PACS line map displayed in Fig. \ref{pacs} shows that the 179\um\, emission is confined along 
the L1448-C outflow, with emission peaks roughly located at the positions of 
shocked spots previously identified through CO and SiO observations;
these are named, following the nomenclature of Bachiller et al. (1990), as R1 to R4 and B1
to B3, for the red-shifted and blue-shifted lobes, respectively. 
The strongest peak is observed towards the central position, where two IR sources,
L1448-C(S) and C(N), are located (J{\o}rgensen et al. 2006). Another strong emission peak is also
observed in the terminal part of the red-shifted lobe (knot R4).
To the north, three different protostellar sources are present, resolved by mm 
interferometric observations and called 
A, B and W following Looney et al. (2000) and Kwon et al. (2006). 
\wat\, emission ends abruptly at the position of L1449 N(A) and N(B), 
while a lane of water in absorption is seen towards L1448 N(W). 
Noticeably, and 
in contrast to what is observed at
the central position, no emission
is associated with any of the three L1448 N sources. 
Given the low spectral resolution
of the PACS observations, there could be a mixture of emission and 
absorption beyond the B2 knot causing 
a near cancellation of the emission.
\begin{figure}
\centering
\includegraphics[angle=0,width=7cm]{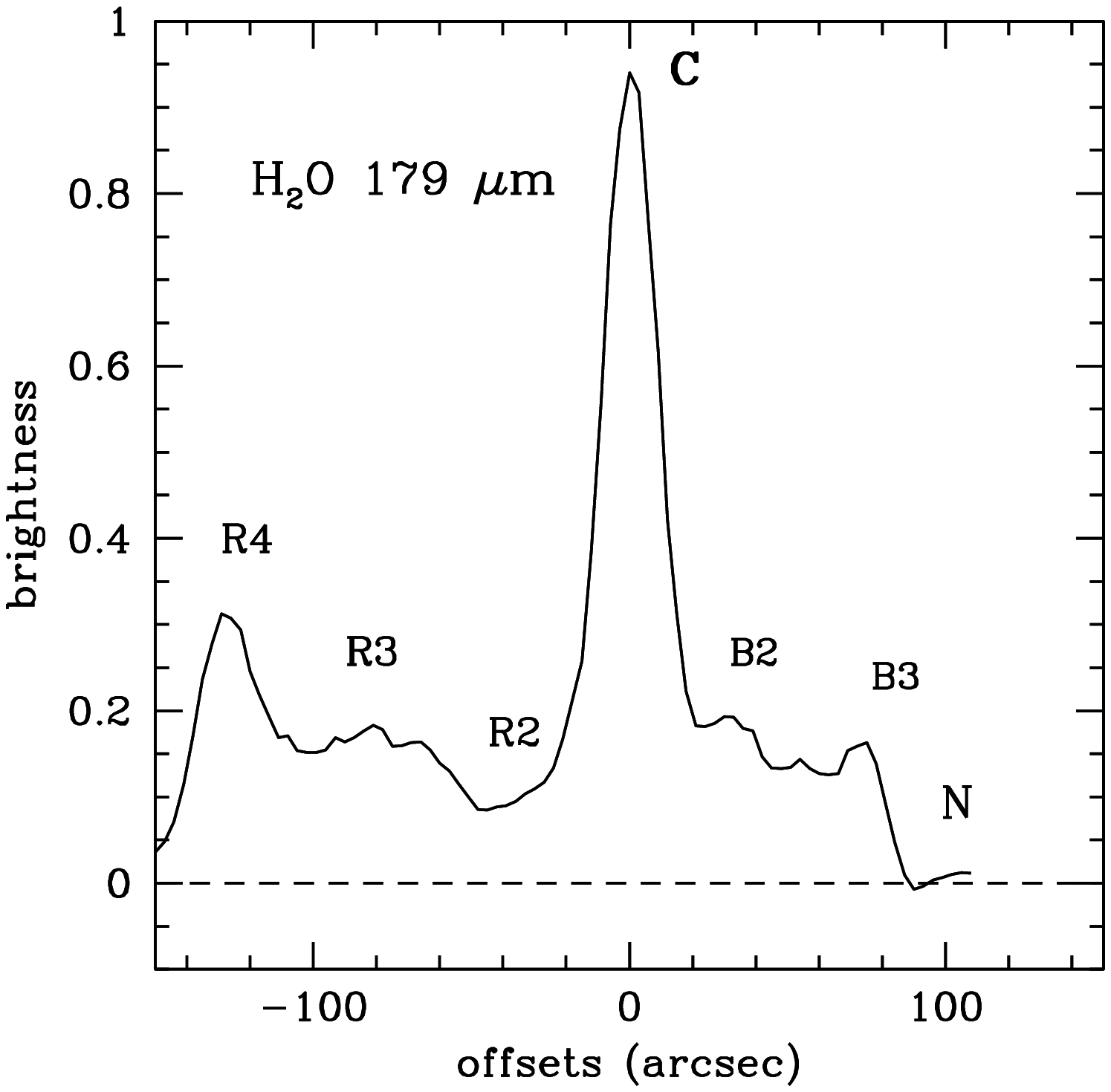}
\includegraphics[angle=0,width=7cm]{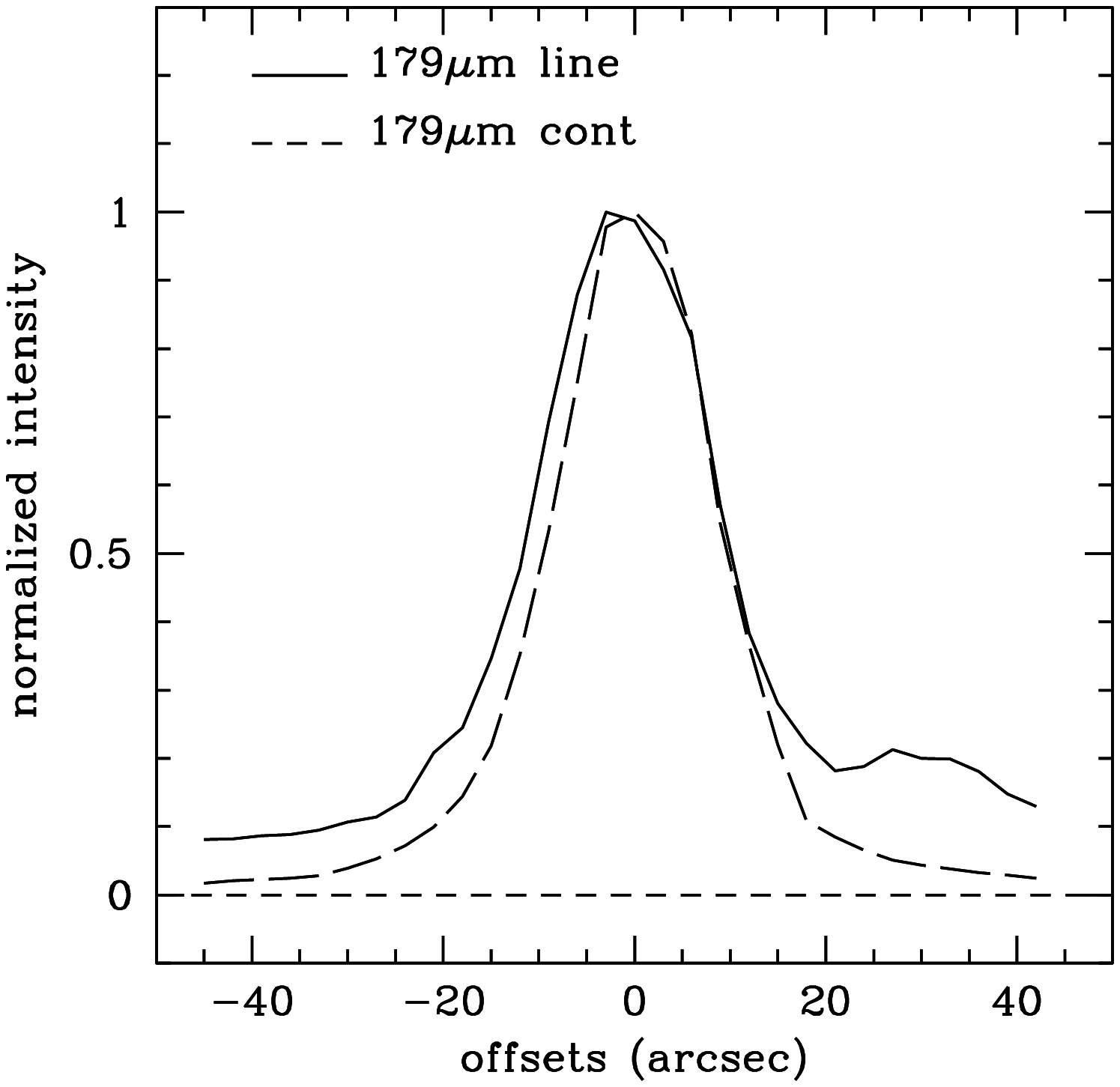}
\caption{{\bf Upper panel:} \wat\,179\um\, intensity cut along the L1448 flow, 
at PA 164$\degr$. The normalized intensity is integrated 
over a width of 30$\arcsec$ perpendicular to the cut.  
The location of the C(N) source is taken as 
the reference for the offsets.
{\bf Bottom panel:} normalized intensities of the 179\um\, \wat\, line and continuum,
along the same cut direction as above.
}

\label{int_cut}%

\end{figure}

Figure \ref{int_cut} shows the line brightness profile along the flow, 
obtained by integrating the intensity in a region with a
width of about 30$^{\prime\prime}$ perpendicular to the outflow axis. 
This plot shows the relative intensities of the 
different emission peaks, indicating that the emission is extended but clumpy. 

In the same figure, bottom panel, an enlargement around the L1448-C source
is displayed, where the 179\um\, line and continuum 
intensity profiles are compared.
In the continuum, the two sources C(S) and C(N) are not spatially resolved: the FWHM
of the spatial profile, when fitted with a gaussian, is $\sim$ 18$^{\prime\prime}$,
which means a deconvolved size of $\sim$ 13$^{\prime\prime}$, assuming both the beam
and the emitting region to be gaussian. For comparison, the two sources have
a separation of about 8$^{\prime\prime}$. The line emission
appears slightly extended with respect to the continuum, with a deconvolved
FWHM of the order of 15$^{\prime\prime}$. This is roughly similar to the extension 
of the inner EHV SiO jet as observed, for example, by
Guilloteau et al. (1992) and 
Hirano et al. (2010) (their knots B1/R1). A more detailed description of the water morphology
in comparison with other tracers in the regions around the C and N sources is
given in Appendix A.

Fig. \ref{CO_HIFI}, right panel, presents a comparison 
of the 179\um\, emission, shown as a grayscale image, with the integrated
\wat\, 557 GHz emission, displayed by separate contours for the blue- and red-shifted emission. 
The two lines 
show a similar morphology
compatible with the much lower spatial resolution of the HIFI spectra. 
Indeed,  as in the PACS data, bright emission
is observed towards the central L1448-C position and the southern, red-shifted
outflow lobe, while no water emission is detected north of the L1448-N
position. The complete set of HIFI data are 
presented in Figure B.1, where all the spectra 
are presented in a regular grid.
The comparison between the
water and CO peaks can be directly 
visualized in Fig. \ref{CO_HIFI}, left panel, where the 179\um\, line map is 
overlaid with contours of the CO(3-2) emission, separated into the blue- and red-shifted gas.
Along the southern outflow, the CO emission is systematically shifted with respect
to \wat. In the northern region, CO extends farthest to the north,  where 
the outflow from the C(N) source is confused with a second 
outflow (at PA roughly 110$\degr$) 
emerging from the N(B) source (Bachiller et al. 1990,
see next subsection). Therefore, although water 
roughly follows the direction
of the CO outflow, there is not a strict correlation between individual emission
peaks.    


\subsection{\wat\, kinematics}

HIFI 557 GHz line profiles at 
selected positions along the outflows
are shown in Fig. \ref{profiles}, 
and compared with the CO(3-2) line.
To make the comparison independent 
of beam filling effects, the CO(3-2) 
lines have been extracted from a map
convolved to the same spatial resolution of the \wat\,557 GHz 
observations. All the lines show narrow absorption at the systemic velocity, 
due to foreground gas. 
The 557 GHz line always traces the same range of velocity as CO.
Maximum velocities up to 50 \kms\, are detected in all positions, while velocities
reaching up to +80\kms\, are observed at the position of the L1448-C source; 
here EHV gas in the form of a separate emission component
 is clearly detected (Kristensen et al. 2011, see
also Fig. 8).
Despite tracing the same velocity range, the \wat\, 
line profiles are different from those of CO,
as already shown in other studies (Santangelo et al. 2012; Kristensen et al. 2011): 
most of the CO emission is localized at low velocity (V$_r  - V_{LSR} \la \pm$ 10 \kms), 
while the bulk of the water emission occurs at intermediate velocities
(V$_r - V_{LSR} \sim$ 5-30 \kms). 

A detailed view of the \wat\, and CO emission spatial distribution, as a function 
of velocity, can be visualized in the velocity channel maps presented in Fig. B.2.
Fig. \ref{channels} shows the maps of the two emissions integrated
in three representative velocity intervals, 
corresponding to the low ($\pm 1-10$\kms), intermediate ($\pm 11-45$\kms)
and high ($\pm 46-86$\kms) velocity gas. 
Emission at the low and intermediate 
velocities is detected all along the outflow, with the exception, as already 
noted from the PACS map,
of the region north-west of the L1448-N sources, where water is absent while 
CO is detected. 
Fig. \ref{channels} also shows that both the \wat\, and CO in the high 
velocity range ($\ga$ 50\kms)
 are not confined at the central source position, but extend between 
 $-$100 and +50 arcsec from L1448-C. 
If we look at the individual spectra shown in 
 Fig. \ref{profiles}, 
we see that in the CO profiles this EHV gas always appears as a separate 'bullet' 
emission superimposed on the line wing at lower velocity (e.g. Bachiller et al. 1990).
These EHV bullets are physically associated with the highly collimated molecular jet displaced 
along the outflow axis (e.g. Hirano et al. 2010). Water emission kinematically associated 
with these bullets is clearly detected only towards the central L1448-C region 
(see Fig. \ref{mm_sio}) and has been discussed 
in Kristensen et al. (2011). 
Although EHV emission is also detected at greater
distances from the source in CO, 
this emission does not appear as a separate bullet component in the individual \wat\, spectra, 
but rather as an extension of the low velocity component wing. 
Hence the contribution of the EHV gas to the total \wat\, line emission is  
smaller than in the case of CO. This will be discussed further in the next section.

\begin{figure*}
\centering
\includegraphics[angle=-90,width=13cm]{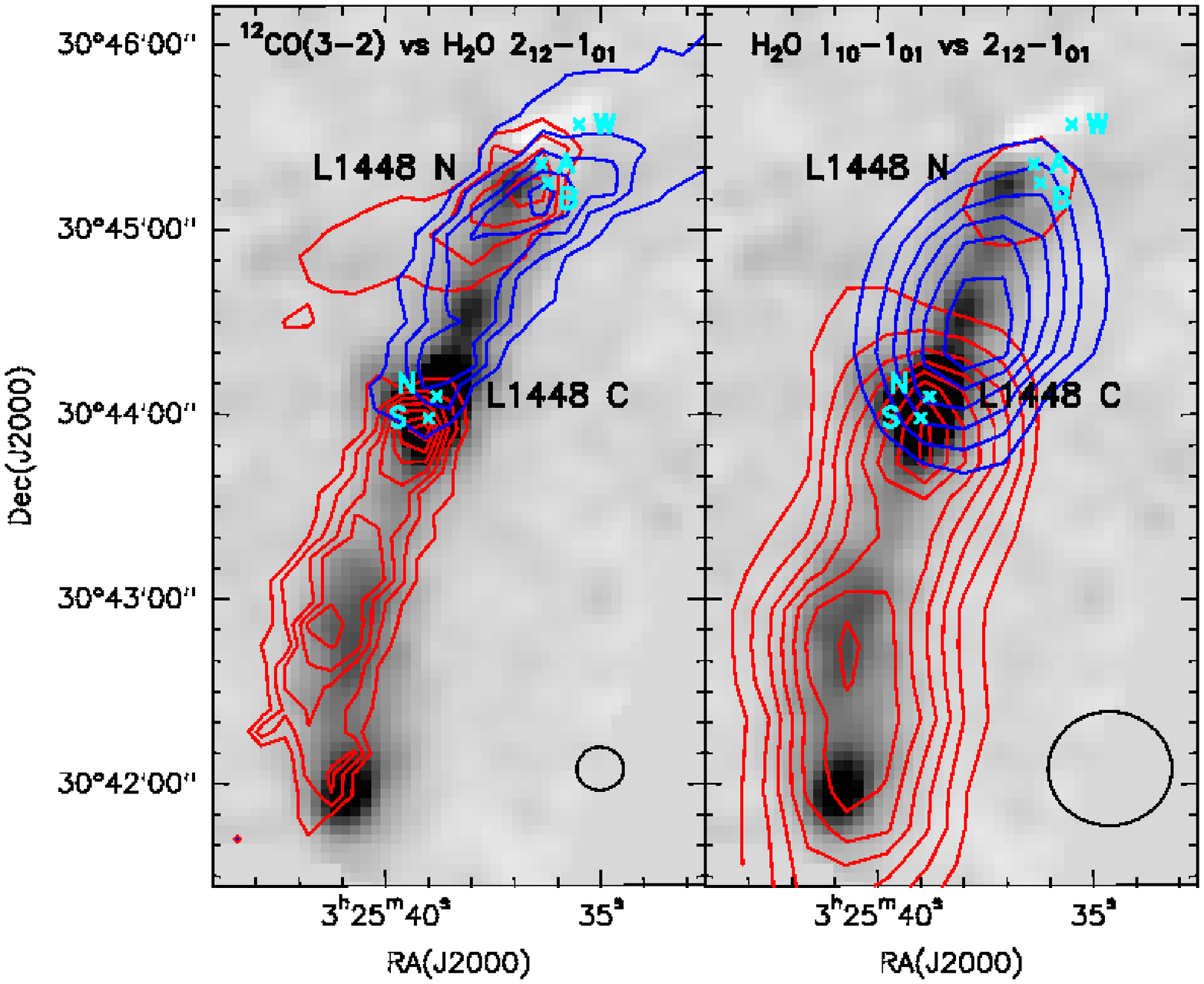}
\caption{Overlay of the JCMT $^{12}$CO(3-2) (\textit{left panel}) and HIFI 
\wat\, 557 GHz (\textit{right panel}) emission on the \wat\, 179\um\, map. The blue and red contours
represent emission integrated in the velocity ranges ($-$100,$+$4) and ($+$6,$+$100) \kms,
respectively. Contours are drawn from 25 to 200 K\,\kms, with steps of 20 K\,\kms,
for the CO(3-2) map, and from 2 to 12 K\,\kms, with steps of 1.2 K\,\kms,
for the \wat\,557 GHz map.
The HPBW of 14$\arcsec$ (JCMT) and 38$\arcsec$ (HIFI) is indicated.
     }
\label{CO_HIFI}%
\end{figure*}

\begin{figure*}
\centering
\includegraphics[angle=0,width=20cm]{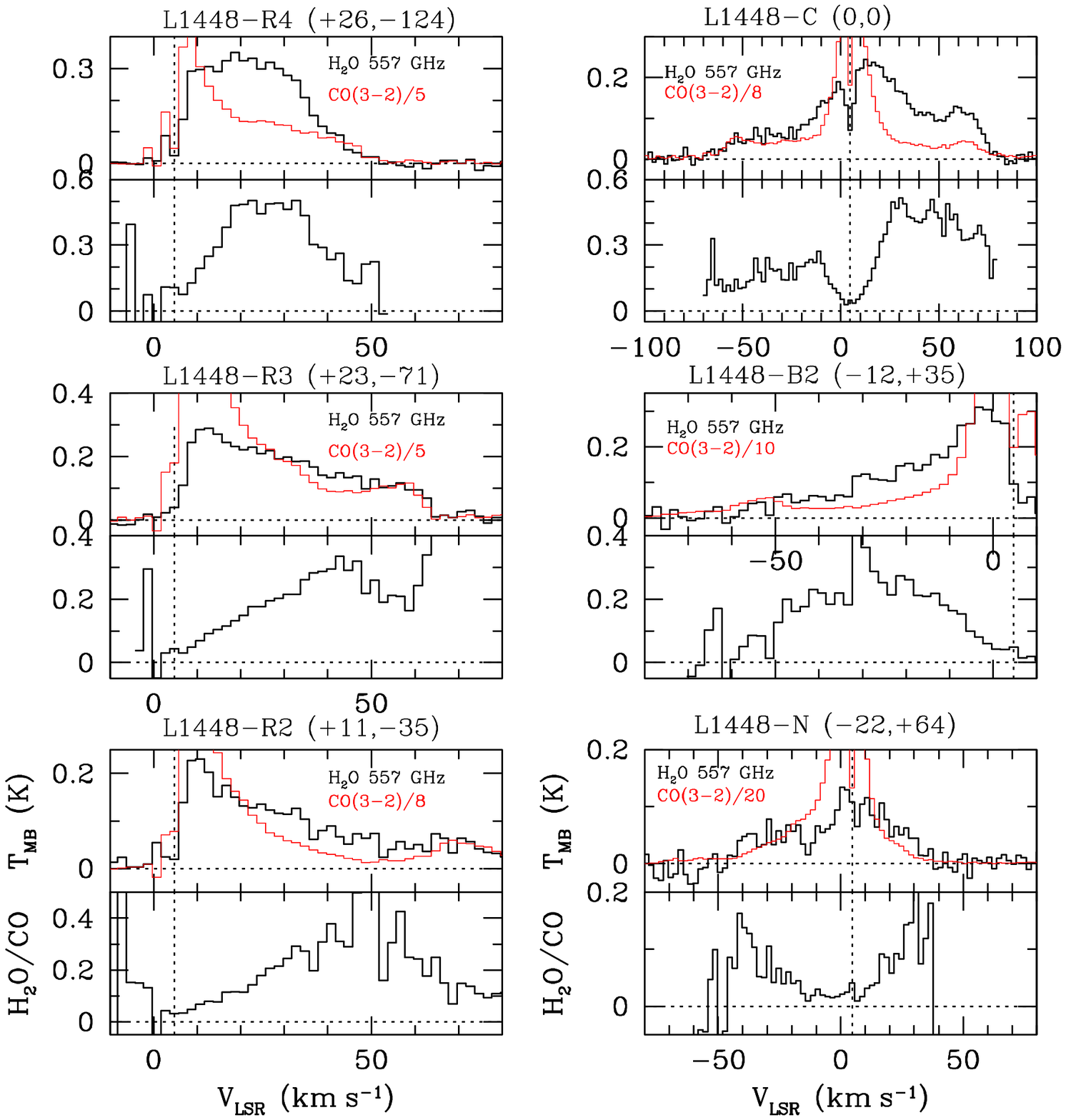}
\caption{\wat\, 1$_{10}$-1$_{01}$ (black) and CO(3-2) (red) line profiles at selected
outflow positions. CO line profiles, convolved 
to the \wat\, spatial resolution, 
 have been scaled to match the \wat\, line wings and the corresponding scaling factor is indicated in each plot. 
At the bottom of each spectrum, the relative \wat/CO intensity ratio is plotted as a function of velocity.}
\label{profiles}%
\end{figure*}

\subsection{\wat-to-CO ratio vs velocity}

As seen in the previous section, the water and CO profiles look different, and
thus their ratio varies significantly with velocity 
as illustrated in Fig \ref{profiles}.  At all the selected outflow
positions, the \wat\, 1$_{10}$-1$_{01}$/CO(3-2) ratio increases with velocity up 
to $v \la$ 20 \kms: 
this is a trend that was identified previously in all sources observed
in the 557 GHz line by \textit{SWAS} and \textit{Herschel} (Franklin et al. 2008; and Kristensen et al. 2012).
Given the high S/N reached in our observations at high velocity, we can now see that beyond
20 \kms\, the ratio reaches a plateau, and
then decreases again at the highest velocities. 
Variations of the \wat/CO line ratio could be due both to variations
in the physical conditions with velocity and/or to abundance
variations. In addition,
at velocities close to the ambient
velocity, a different 
degree of absorption of the two lines by the cold gas 
may influence this ratio. The increase in the \wat/CO ratio as a function of velocity
has been so far interpreted as an increase 
of the \wat\, abundance at high speeds; assuming the same temperature and density 
 conditions for the two lines, Franklin et al. (2008) derived an \wat\, abundance 
 in the gas with $v_{max} \sim 20$ \kms\, an order of magnitude higher 
than that in the low velocity gas.
This conclusion, however, was based on an erroneous assumption, since 
 different physical conditions pertain to CO and \wat; 
moreover, the physical conditions change with
velocity, as shown in, e.g., Santangelo et al. (2012), Vasta et al. (2012) and 
Lefloch et al. (2010).
 
Furthermore, the decrease of the \wat/CO line ratio at velocities larger than $\sim$ 30-40 \kms\,
contradicts the conclusion that a larger water abundance is always associated with the
gas at the highest velocity.
The drop in 
the \wat/CO line ratio roughly coincides with the velocity range of the EHV bullet emission,
indicating that a critical change in the physical and/or
chemical conditions occurs in the bullets with respect to the 'standard' wing emission.
Tafalla et al. (2010) 
studied the chemical composition of the EHV gas in L1448, 
comparing it with the gas responsible for the wing emission, and found
significant differences between these two 
components. 
They found, in particular, that the EHV gas is relatively rich in 
O-bearing species and poor in C-bearing molecules compared to the wing regime.  
Thus the 
observed drop in the \wat\,1$_{10}$-1$_{01}$/CO(3-2) ratio in the EHV regime is more 
easily understood
if the water-emitting EHV gas has a lower temperature and/or density 
that the gas responsible for the wing emission. Kristensen et al. (2011) 
compared the excitation conditions 
for water in the EHV gas with those responsible for the wing emission towards the central position, 
but were unable
to identify significant differences in the two regimes. Santangelo et al. (2012), 
on the other hand, found that 
at the L1448-R4 position the high velocity \wat\, emission 
is associated with gas at a density about an order of magnitude lower than 
that of the gas responsible for the low velocity emission. 

\subsection{SiO and \wat}

\begin{figure}
\centering
\includegraphics[angle=0,width=9cm]{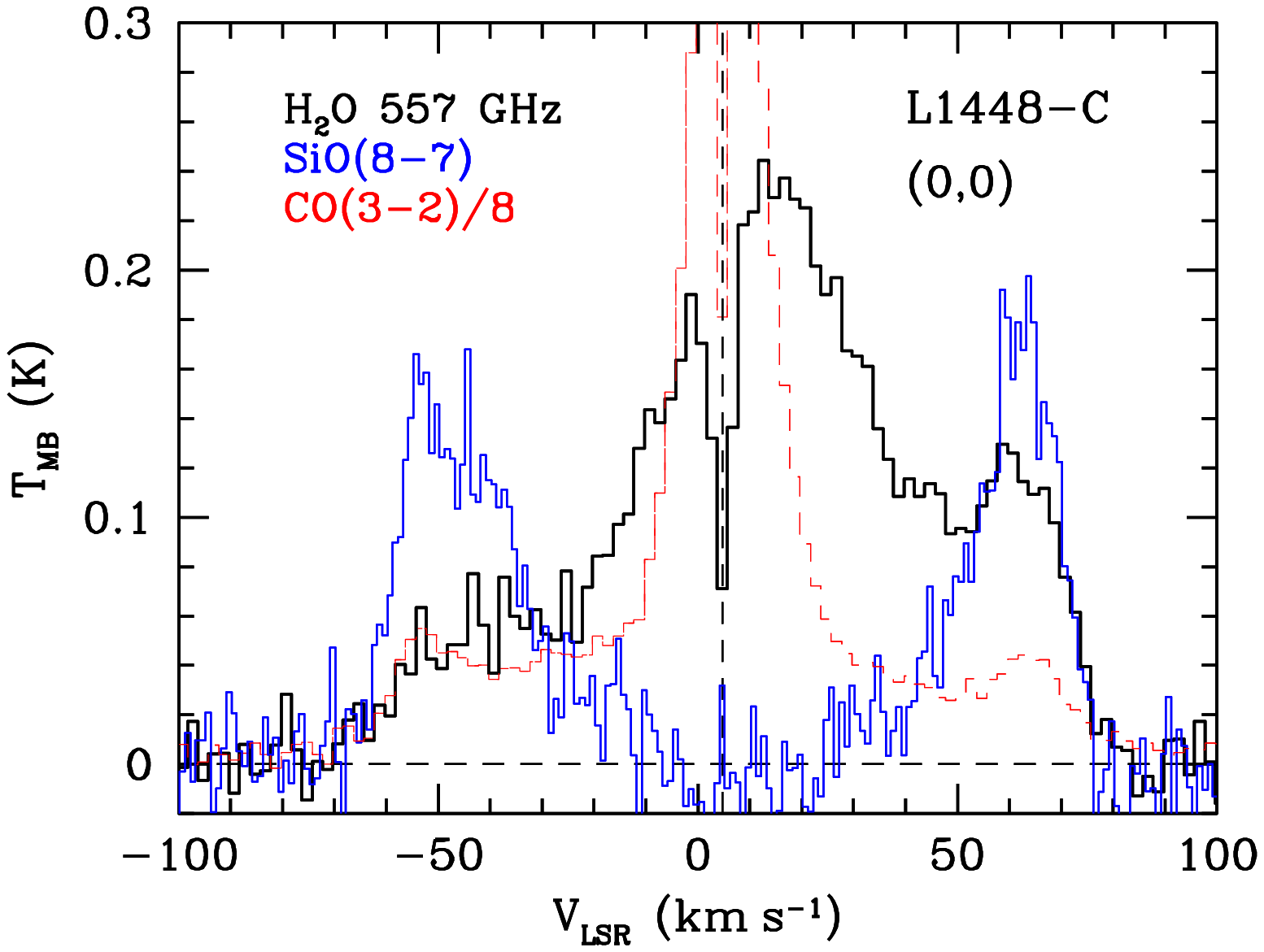}
\caption{Overlay between \wat\, 1$_{10}$-1$_{01}$ (black), SiO(8-7) (blue) and
CO(3-2) (red) towards the L1448-C position. 
}
\label{mm_sio}%
\end{figure}

\begin{figure*}
\centering
\includegraphics[angle=-90,width=15cm]{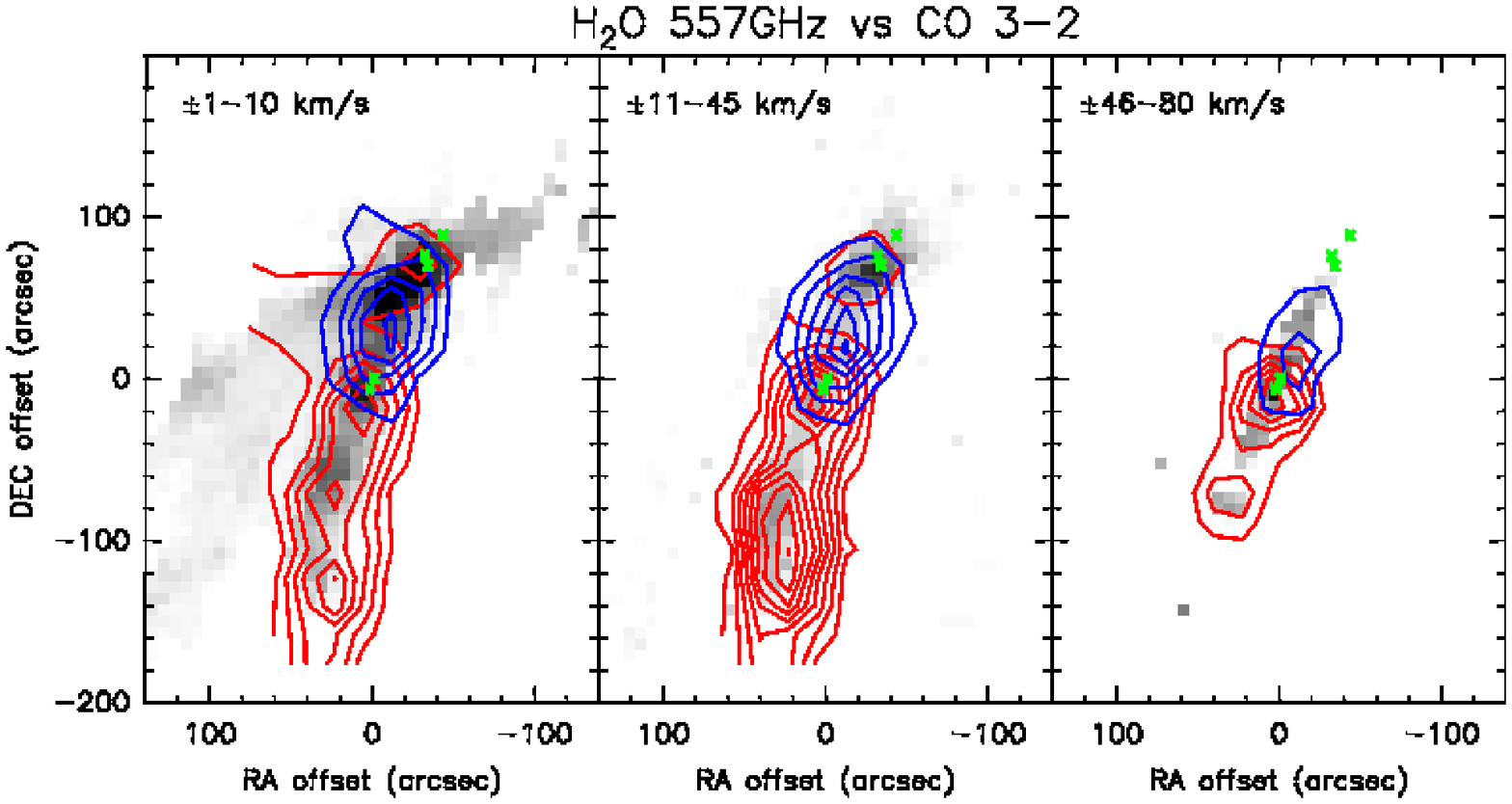}
\caption{Contours of the \wat\, 1$_{10}$-1$_{01}$ emission integrated in three different
velocity intervals, superimposed onto the CO(3-2) emission in the same
velocity bins (gray scale). The velocity ranges with respect to the systemic velocity 
of V$_{LSR}$ = 4.7\kms, are shown in the upper side of each panel. Green crosses
mark the positions of the different sources (see Fig.1). 
     }
\label{channels}%
\end{figure*}

The SiO(8-7) map, obtained together with the CO(3-2) observations, 
only shows emission close to the central position, where it can be associated with the L1448-C 
molecular jet (Hirano et al. 2010). In fact, while SiO emission from the (1-0) and (2-1)
 transitions is observed along the entire molecular outflow, peaking at the different 
 clump positions (Bachiller et al. 1991, Dutrey et al. 1997), lines at higher excitation are 
observed only towards the highly collimated micro-jet
(Bachiller et al. 1991, Nisini et al. 2007). 
The comparison of the SiO and \wat\, emissions (see Fig. \ref{mm_sio})shows that their profiles
are strikingly different. The EHV bullets are 
more prominent in SiO than in \wat: conversely, no SiO is associated
with the strong intermediate velocity broad \wat\, emission peaking 
around the ambient velocity. The association of SiO emission with the EHV
collimated jet is a well-known feature characteristic of several class 0 sources
(e.g. Hirano et al. 2006; Codella et al. 2007); it has been suggested that SiO in the jet 
is either directly synthesized 
in the dust-free jet acceleration region (Glassgold et al. 1991; Panoglou et al. 2011)
or originates in shocked ambient material where silicon is released 
into the gas phase by
the disruption of grain cores (e.g. Gusdorf et al. 2008).

An origin
in the primary jet has been recently supported by both interferometric 
observations in the HH212 object (Cabrit et al. 2007) and by the molecular survey
conducted on the L1448-R2 EHV bullet by Tafalla et al. (2010), who
showed that the bullets possess a peculiar chemistry with respect to
 the standard outflow wing emission, suggesting an origin different from shocks.
The fact that SiO(8-7) is more prominent than \wat\, 557 GHz in the EHV bullets 
may either be an excitation effect or 
the result of an enhanced SiO/\wat\, 
abundance ratio, or both. 
The excitation of SiO in the EHV bullets has been 
studied by Nisini et al. (2007), who found that the 
SiO-emitting
gas has a density
$\sim$ 10$^6$ \cmt\, and $T_{kin} \ga$ 300 K. Kristensen et al. (2011) found
that similar conditions may be consistent also with the water emission in the bullets,
suggesting that SiO and \wat\, are excited in the same gas. 
With this assumption, the observed \wat\,557 GHz/ SiO(8--7) intensity ratio
in the bullets implies a \wat/SiO abundance ratio of $\sim$ 10. 
Shock models that take 
account of the erosion of Si from grain cores and mantles
predict this ratio to be of the order of 10$^3$ or more, depending on the 
shock velocity (Gusdorf et al. 2008; Jim\'enez-Serra et al. 2008). On the other hand
a \wat/SiO ratio of about 10 is predicted by the wind model of Glassgold et al.(1991)
where \wat\, and SiO are formed in dust-free gas directly ejected from the
protostar, provided that the mass loss rate of the spherical wind 
is $>$ 10$^{-5}$ M$_\odot$\,yr$^{-1}$.
Only such high mass loss rates yield a density
at the wind base that is high enough to permit efficient SiO synthesis
through gas-phase reactions. Indeed, timescales for SiO production 
 are rather low, i.e. they stay below 10$^2$ yr, for $T > 400 K$, 
only if the gas density is $\ga$10$^7$-10$^8$\cmt.
Dionatos et al. (2009) measured for
the L1448 jet a molecular mass flux rate of $\sim$ 10$^{-7}$ M$_\odot$\,yr$^{-1}$:
for this low mass loss value, the model by Glassgold et al. (1991) predicts
a negligible abundance of both SiO and \wat. However, given the high collimation
of the L1448 jet, the mass loss rate values are not directly comparable 
and certainly the possibility that the two molecules trace the primary
jet cannot be ruled out. In this respect, initial results presented in Panoglou
et al. (2011) for the molecular survival in disk-winds seem promising, 
predicting that a significant fraction of water is synthesized in 
jets from class 0 sources having a mass accretion rate of 5\,10$^{-6}$ M$_\odot$\,yr$^{-1}$,
implying a mass flux rate of the order of that measured in L1448.

Finally, we note that the timescales to increase the water abundance
to values X(\wat)$>$10$^{-5}$ in a gas with $T >$ 400 K are of the order of 100 yr (Bergin et al.
1998), which match well
the dynamical timescale for the L1448 jet propagation
of the order of $\sim$ 150 yr (Hirano et al. 2010).

With regards to the broad \wat\, emission at intermediate velocities,
Kristensen et al. (2011) suggested an origin in shocks 
caused by the interaction between the outflow and the envelope. 
Such shocks would be expected to produce significant SiO emission, 
since the disruption of grain cores occurs at shock speeds $\ga$ 25-30 \kms 
(Jim\'enez-Serra et al. 2008; Gusdorf et al. 2008). The efficiency of
sputtering and grain-grain collisions, however, depends on the type
of grains involved and on the total density: for large grains, sputtering 
can be significantly inhibited for n(\h)$\ga$ 10$^6$ \cmt, due
to the decrease, at such densities, of the relative velocity between 
grains and neutral species (Caselli et al. 1997). 
In fact, the observed SiO(8-7) emission gradually rises from the ambient velocity 
up to the EHV regime, behaviour 
which could suggest a progressive enhancement of the 
SiO abundance 
moving from the regime of high density and low velocity to that of 
low density and high velocity;
 the water, on the other hand, can be 
efficiently produced even at low shock speeds and high densities 
from sputtering of icy grain mantles, which would explain the different
behavior of the two species in the intermediate velocity regime.
However, the non-detection towards L1448-C of broad lines 
from other molecules residing on ices, such as CH$_3$OH (Jim\'enez-Serra et al. 2005), 
is indicative of the fact that the gas/grain chemistry can  
indeed be more complex than normally assumed.

\begin{figure}
\centering
\includegraphics[angle=0,width=10cm]{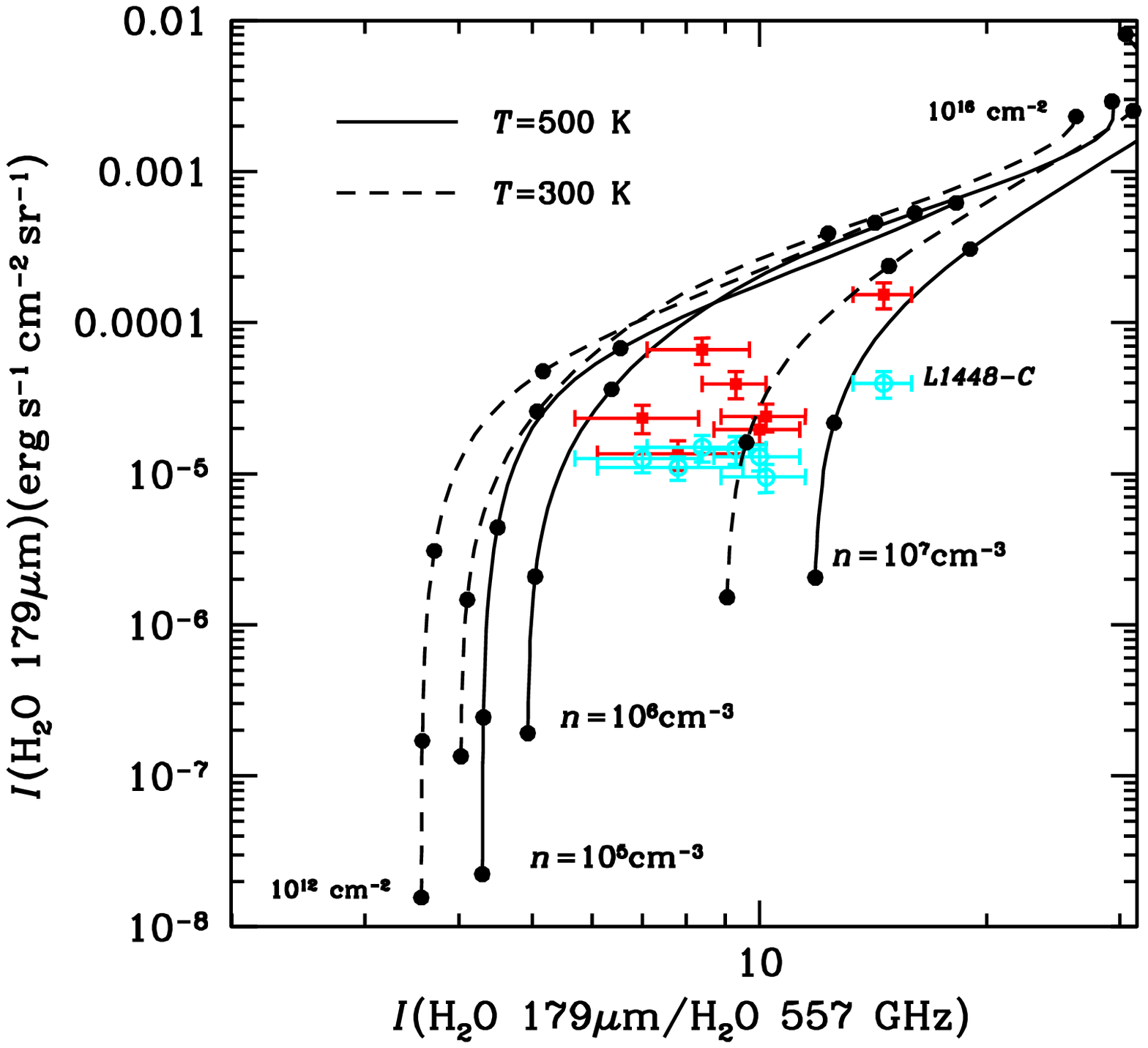}
\caption{\wat 179\um /557GHz ratio vs
   the \wat 179\um\, intensity (expressed in erg\,s$^{-1}$\,cm$^{-2}$\,sr$^{-1}$). 
   Theoretical LVG predictions are
   plotted for volume densities of 10$^{5}$, 10$^{6}$ and 10$^{7}$ \cmt,
   and for kinetic temperatures of 300 K (dashed line) and 500 K (full line). 
   Along each curve the $N$(o-\wat) column density varies from 10$^{12}$ 
   to 10$^{16}$ \cmd\, in steps of a factor 10. 
Filled red squares refer to intensities measured
   in a beam of 38\arcsec, equal to the beam 
size of the 557 GHz observations. 
   Open cyan circles plot the observed values with the y-axis indicating the {\it unconvolved} PACS intensity.
  The data point for L1448-C is labeled.
     }
\label{diag179}%
\end{figure}

\begin{figure}
\centering
\includegraphics[angle=0,width=10cm]{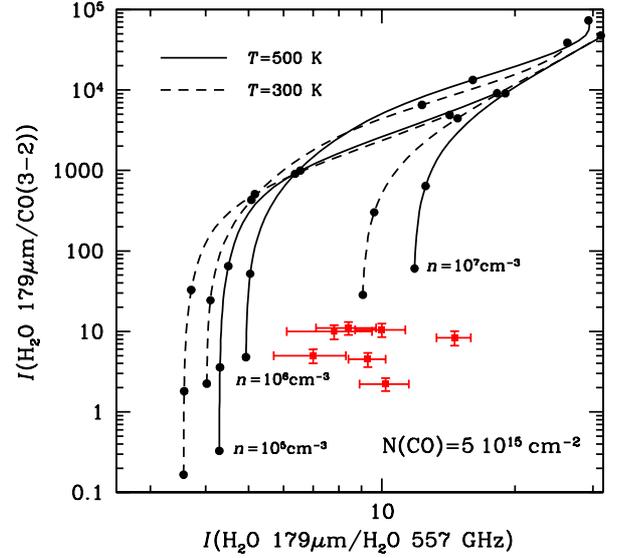}
\caption{Diagnostic diagram employing the \wat\,179\um/557 GHz vs the \wat\,179\um/CO(3--2)
intensity ratios. Theoretical LVG predictions are plotted for the same parameters
as in Fig. \ref{diag179}. The plot assumes a fixed $N$(CO) column density of 5\,10$^{15}$ \cmd
(see text for details). Data points refer to intensities measured
   in a beam of 38\arcsec, equal to the beam of the 557 GHz observations. 
     }
\label{diag_co}%
\end{figure}


\section{H$_2$O physical conditions and abundances}

From the relative and absolute intensities of the observed \wat\, lines,
it is possible to derive spatially- and spectrally-averaged information about 
their excitation conditions. 
For this purpose, we have convolved the
PACS line map at the HIFI 557 GHz resolution (i.e. $38\arcsec$),
and we have integrated the HIFI spectra over velocity,
in order to compare line intensities for the same spatial and
spectral regions. In Table 1, we report these intensities, as measured
at different positions along the outflow, 
corresponding roughly
to the water intensity peaks.  

In Fig. \ref{diag179}, the 179\um\, intensity is plotted as 
a function of the 179\um /557GHz ratio. Here the observed values are confronted with
predictions obtained using the RADEX code (van der Tak et al. 2007)
that we have run using the Large Velocity Gradient (LVG) approximation in plane-parallel 
geometry. 
Based on the analysis presented in Santangelo et al. (2012), Vasta et al. (2012)
and Bijerkeli et al. (2012), we assumed that the kinetic temperature ($T_{kin}$) 
of the gas traced by the observed \wat\, transitions is in the range 300-500 K,
similar to 
that derived by Giannini et al. (2011) from \textit{Spitzer}
observations of the low-lying H$_2$ transitions.
We then explored hydrogen densities ($n_{H_2}$)
in the range 10$^5$-10$^7$ \cmt\, and o-H$_2$O column densities in the range
10$^{12}$-10$^{16}$\cmd. A linewidth of 30\kms\, 
was adopted, representing the
typical FWHM of the observed 557 GHz line.

In Fig. \ref{diag179} \wat\, 179\um\, intensities, convolved to the 557 GHz resolution,
are indicated as filled (red) squares, while open (cyan) squares indicate
the unconvolved PACS intensities. 
The convolved and unconvolved intensities differ by factors between 1.2 and 5, 
reflecting the extended but clumpy nature of the 179\um\, emission as
shown in Fig. \ref{int_cut}.
Assuming that the unconvolved intensities are not further diluted within
the PACS beam, Fig. \ref{diag179}, suggests that the density of the gas responsible
for the \wat\, emission is in the range $\sim$ 10$^6$-10$^7$ \cmt\,
while the \wat\, column density is $\ga$ 10$^{13}$ \cmd.

This result is consistent with the work of Santangelo et al. (2012), who analysed  
\textit{Herschel}-HIFI spectra of several \wat\, lines gathered towards the L1448-R4 and B2 
positions, concluding that the water emission in these positions arises  
from a gas at T$\sim$ 400-600 K and density of the order of 1-5\,10$^6$\cmt. 
Models with densities lower than 10$^6$\cmt\, were not able to fit
all the lines observed with HIFI and would require 
high column densities that are inconsistent
with the upper limit on the H$_2^{18}$O observed in the L1448-R4 position.
The conditions we derived
are also consistent with the results 
obtained by Tafalla et al. (2012) from an analysis of
the o-\wat\, 1$_{10}$-1$_{01}$ and 2$_{12}$-1$_{01}$ 
emissions in a large sample of shocked spots.

In Fig. \ref{diag_co} the \wat\, data are also compared with the CO(3-2) line intensity,
which has been convolved to the HIFI resolution.
The expected 179\um/CO(3-2)  ratio has been computed for $N({\rm CO})$ = 5\,10$^{15}$ \cmd:
this value assumes an average $N({\rm H_2})$ column density of 5\,10$^{19}$ \cmd\,
along the flow (consistent with the column density estimated in Giannini et al.\ 2011)
and a CO abundance of 10$^{-4}$. 
The figure shows that the observed data are reproduced by a single gas
component only for the case of $n({\rm H_2})$ $\sim$ 10$^7$ \cmt\, and 
$N({\rm H_2O})$ $<$ 10$^{12}$ \cmd. As shown above, such conditions 
are not consistent with intensities of the 179\um\, emission.
We also note that the assumed $N({\rm CO})$ is only an upper limit 
on the beam-diluted CO column density in the assumed 38$\arcsec$ aperture, since the 
$N({\rm H_2})$ has been estimated over a beam of size $\sim$ 10$\arcsec$.
To be consistent with the observed ratios, a CO column
density one or two order of magnitudes larger than 
that assumed here would be required,
clearly inconsistent with the H$_2$ observations. 
This is further and more quantitative evidence that the CO(3-2) and the water emission originate
from gas components with different excitation conditions. 
The low-$J$ CO is likely
related to the cold entrained gas and not directly 
associated with the high temperature
shocked gas. Assuming that the CO emission 
originates in gas at a temperature of the 
order of 100 K , and with a density of 10$^4$\cmt (e.g. van Kempen et al. 2009),
the contribution  
of the \wat\, 557 GHz emission from this gas would be negligible (i.e. 1/10 of the
observed beam diluted intensity), even assuming a \wat/CO abundance $\sim$1. 
Assuming $T_{kin} \sim$ 500 K and $n_{H_2}\sim$ 5\,10$^6$ \cmt, o-\wat\, 
beam-diluted PACS column densities 
are constrained to be of the order of 10$^{13}$ \cmd\, along the flow, while a higher value, 
of the order of 10$^{14}$ \cmd\, is inferred towards the central position (see Table 1). 

To estimate the corresponding \wat\, abundance, we derive 
$N({\rm H_2})$ from the \textit{Spitzer} spectral image of the H$_2$ S(1) 17\um\, line
(Neufeld et al. 2009; Giannini et al. 2011), 
on the assumption that it originates in the same gas as seen in H$_2$O by PACS. 
For this purpose, the H$_2$ 17\um\, image was convolved  
to the 12\arcsec resolution of the PACS map, and the beam-averaged $N({\rm H_2})$ 
was determined 
assuming the line to be in LTE with T$_{kin}$ = 500 K.
Values of $N({\rm H_2})$ of the order of 0.7--6\,10$^{19}$ \cmd\, are derived
for the different positions.
Consequently, the water abundance along the outflow is relatively constant
with values $\sim$0.5--1$\times$ 10$^{-6}$ (see Table 1). 
For the on-source
position, a larger value of $\sim$ 10$^{-5}$ is found; however, 
the large on-source extinction might cause the $N({\rm H_2})$ column density to be
underestimated, and consequently the X(\wat) to be overestimated. 
If we consider as a reference the H$_2$ S(0) 28\um\, emission for deriving the 
$N({\rm H_2})$ towards the central position, we obtain a value of $\sim$ 5\,10$^{20}$ \cmd\, ,
which would imply X(\wat)$\sim$ 10$^{-6}$. 
In this case, we should consider this value as a lower limit, since the S(0)
emission is likely dominated by gas colder than that giving rise
to the water emission observed here. 

%
\begin{table*}
\caption{Line intensities$^*$ and H$_2$O abundances}             
\label{table:1}      
\begin{tabular}{ccccccccc}        
\hline\hline                 
Position & 
\multicolumn{3}{c}{$\int{T_{mb}dv}$ (K\,\kms)} &
\multicolumn{3}{c}{$I$(10$^{-6}$erg\,cm$^{-2}$\,s$^{-1}$\,sr$^{-1}$)} &
$N$(o-H$_{2}$O)$^b$ & X(\wat)$^c$
\\
\cline{2-4} \cline{5-7} \\
(\arcsec,\arcsec)\tablefootmark{a} 
&\wat 
& \wat
& CO 
& \wat
& \wat
& CO  
&10$^{13}$\cmd 
& 10$^{-6}$\\
&
$1_{10}$-$1_{01}$ &
$2_{21}$-$1_{01}$ &
$J=$3--2 &
$1_{10}$-$1_{01}$ &
$2_{21}$-$1_{01}$ &
$J=$3--2 &
 &\\
\hline                        
($+$26.0,$-$124.1) &10.3 & 3.2 & 32.9 &1.78 &14.9 & 1.3 &3.0 & ... \\
($+$29.3,$-$98.2)  &7.5  &2.8  & 31.6 &1.30 &12.9 & 1.3 &1.3  & 0.7\\
($+$23.4,$-$71.1)  &10.4 & 2.7 & 61.3 &1.80 &12.6 & 2.5 &1.6 & 0.8\\
($+$11.1,$-$36.4)  &7.9  &2.4  & 26.9 &1.37 &11.0 & 1.1 &1.0  & 2\\
(0,0)              &15.7 &8.5  & 115.6&2.72 &39.6 & 4.7 &9.0 & 1-12.0\\
($-$12.5,$+$34.9)  &9.0  &3.0  & 78.05&1.56 &14.5 & 3.2 &1.6  & 0.8\\
($-$21.7,$+$63.9)  &5.4  &2.1  & 104.3&0.93 &9.49 & 4.2 &1.8 & 0.4\\
\hline                                   
\end{tabular}
\\
\\
\tablefoottext{*}{Intensities measured in a circular area of 
diameter 38\arcsec. Absolute uncertainties
are of the order of 20-30\% for \wat\, measurements and 15\% for CO(3-2).
}
\\
\tablefoottext{a}{Offsets with respect to $\alpha_{2000}$ = 03:25:38.8, 
$\delta_{2000}$ = $+$30:44:04}
\\
\tablefoottext{b}{Derived from the unconvolved $2_{21}$-$1_{01}$ line intensity assuming
$T_{kin}$ = 500 K and $n$(H$_2$) = 5\,10$^6$\cmt. The associated uncertainty is 
within a factor of 5.}
\\
\tablefoottext{c}{With respect to the H$_2$ column densities derived from the S(1) 17\um\,
line and assuming an o/p ratio of 3. Uncertainty on the abundance is within a factor of 10.}
\end{table*}

We note that the derived abundances are sensitive
to the adopted parameters and assumptions. In the  
regime considered here, the water column densities that we derive depend almost linearly on the 
assumed \h\, density, which we estimate to be uncertain by a factor of five.
Changes in the assumed temperature, on the other hand, will affect both the \wat\, 
and the \h\, column densities in a similar fashion, having less impact on the derived \wat\, abundance. 

\section{Origin of the observed emission}

Our analysis of the H$_2$O 1$_{10}$-1$_{01}$ and 2$_{12}$-1$_{01}$ lines
suggests that the gas responsible for the bulk of the \wat\, emission is warm, 
with T$_{kin} \sim$ 300- 500 K, and 
very dense, with $n_{H_2}\sim$ 5\,10$^6$ \cmt.
These parameters, as well as the associated low abundance of $\la$\,10$^{-6}$,
seem to be typical of the excitation traced by the two \wat\, transitions, 
since similar physical conditions have been 
derived for other outflow
positions by several authors 
(e.g. Bjerkeli et al. 2012, Vasta et al. 2012, Tafalla et al. 2012).
These physical conditions, 
along with the observed spatial distribution of the 
179$\mu$m emission, indicate that these \wat\, lines
mainly trace gas which has been heated and compressed 
by shocks, rather than entrained ambient gas. 
This latter possibility was suggested 
by Franklin et al. (2008) on the 
basis of\textit{ SWAS} observations, but assuming 
for the 557 GHz line the same physical parameters as the CO(1-0) line.
Our maps have in addition  
provided evidence that the excitation conditions 
and abundance of water in L1448 are fairly constant at the sampled spatial scales. 
This implies very similar shock properties, which seem not to be affected 
by evolutionary effects on the timescales of outflow propagation.
The only exception is the region immediately 
adjacent to the protostar L1448-C:
here an order of magnitude larger \wat\, column density is found 
relative to the other outflow positions. Part of this on-source emission is associated
with the EHV jet where \wat\, and SiO molecules might be directly synthesized 
in the atomic free protostellar wind (see also Kristensen et al. 2010).

The temperature of few hundred K inferred along the outflow
is much lower than the maximum 
temperature of shocked molecular gas, as traced by \h\, near-IR rovibrational lines
($\sim$ 2000 K), for example. 
Far-IR \wat\, lines at higher excitation 
observed by \textit{ISO} in L1448 indicate the presence of hotter gas 
with T$\ga$ 1000 K (Nisini et al. 1999, 2000), thus suggesting 
a distribution of gas temperatures,
as has been inferred for the \h\, gas. 
PACS observations of several young sources suggest that the presence of
gas components at different temperatures is indeed very common 
(e.g. Herczeg et al. 2011; Karska et al. 2012; Goicoechea et al. 2012)
and that the \wat\, abundance is typically larger in the hotter gas 
(Giannini et al. 2001, Santangelo et al. in preparation).

Considering excitation in a single shock, one can expect that 
different excitation components are associated with
different layers in the post-shock region, and that  
the low-lying \wat\, transitions considered here should trace post-shocked gas layers
where the gas has already cooled down to a few hundred K.
Santangelo et al. 2012  
inferred that the 
ratios of low excitation \wat\, lines
in the L1448-B2 and R4 spots are consistent with non-dissociative J-type shocks, 
a conclusion also supported by the large inferred densities, 
which imply a large
shock compression factor. 
In such shocks, as also in C-shocks, high \wat\, abundances are 
produced in the hot gas, due to the rapid conversion 
of atomic oxygen into water when $T_{kin}$ exceeds $\sim$ 300-400 K 
(Kaufman \& Neufeld 2006; Flower \& Pineau Des For{\^e}ts 2010),
and should be maintained long after 
the gas has cooled down. Hence, we should measure the same high \wat\, abundance
in both the warm and the hot gas, unless the density is so high
to allow a very quick freezeout of gas-phase water onto grain mantles.

For $n_{H_2} \sim$ 10$^6$ \cmt, Bergin et al. (1996) derived 
a timescale
$\sim$ 10$^4$ yr for this process. Typical timescales for J-shock
propagation at a pre-shock density of 10$^4$\cmt\, are less than 
10$^2$ yr and the timescales are not longer than $\sim$ 10$^3$ yr even for C-type shocks; 
hence grain freeze-out will be
of minor importance in reducing the gas-phase water column density in
the still warm regions of the post-shocked gas. A water abundance 
smaller than that expected to result from endothermic gas-phase reactions 
could result if most of the oxygen not in CO is frozen out
in ice mantles in the pre-shock gas. Such a possibility is strongly suggested
by the very low abundance of O$_2$ gas as measured
by \textit{SWAS} and \textit{Odin} in dense molecular clouds, which indicates 
that atomic oxygen could be largely depleted (Goldsmith et al. 2000, Larsson et al. 2007).
However, ice mantles are quickly destroyed by sputtering for shock speeds exceeding
$\sim$ 10-15\,\kms, 
so freeze-out within the pre-shock gas can be of relevance only for
very slow shocks.

A different way to decrease the \wat\, abundance in the post-shocked gas 
could be through photodissociation by a pre-existing FUV field. 
Shock regions located along the outflow cavity wall close to the protostar 
could be directly exposed to the central source FUV field (e.g. Visser et al. 2012).
Far from the source, the only way
to produce a significant FUV field is through fast J-type dissociative shocks. 
This scenario assumes a superposition of two shocks at different
velocities: this is expected, e.g., in jet driven outflows where 
a fast dissociative shock (i.e. the jet shock or Mach disk) decelerates the jet 
and a low speed shock accelerates the ambient medium (e.g. Raga \& Cabrit 1993). 
The effect of FUV photons, generated
by a J-shock, impinging on the region 
behind a non-dissociative shock has been discussed in Snell et al. (2005). Their result
is that these photons are not effective in decreasing the abundance of the
hot \wat\, produced at the shock front, since here the timescales for
\wat\, formation are extremely short. 
However, in the post-shocked cooling region, 
water can be rapidly dissociated and consequently the \wat\, 
abundance decreases significantly from the peak value at the shock front.
Timescales for \wat\, dissociation 
depend on the strength of the FUV field and the degree of FUV shielding (Lockett et al. 1999). 
Direct exposure to a
radiation field with $G_0 > \rm few \times 10$ (where $G_0$ is the intensity of the radiation
field relative to its average interstellar value) returns all the oxygen 
to atomic form very quickly. 
If the field is shielded by an $A_v \sim 1$ mag, timescales for converting \wat\, back
to oxygen are of the order of thousands of years and still compatible
with the outflow dynamical timescale.
Snell et al. determined that the column of post-shocked \wat\, 
behind a C-type shock should scale with the FUV field as 
$\sim 4\times10^{15} (n_0/10^4{\rm cm}^{-3})(v_s/10\,{\rm km\,s^{-1}}) G_0^{-1}$, for a shock of velocity $v_s$ in gas of pre-shock density 10$^4$\cmt. 
Our derived column densities of the order of 10$^{14}$\cmd\, therefore
imply the presence of a FUV field $G_0 \sim 25 (n_0/10^4\,\rm{cm}^{-3})^{-1}(v_s/10\,\rm {km\,s^{-1}})^{-1}\,\rm cm^{-2}$ for such a shock. 
Further modeling will be required to determine the exact properties of a J-shock capable of producing the necessary
FUV field. However, the jet speed (with projected velocity up to 80\kms\, along the line of sight) is 
certainly high enough to drive a J-type shock that emits strongly at FUV wavelengths.
The presence of a dissociative shock giving rise to ionizing photons is
suggested, in at least specific positions, by the 
detection of [Fe II] emission along the L1448 outflow by Neufeld et al. (2009)
and of OH emission towards the B2 clump (Santangelo et al. 2012). Shocks
close to the sources could be instead directly exposed to the source FUV field expanding
in the envelope cavity, whose presence is 
revealed by the scattered light
emission detected in the \textit{Spitzer} IRAC images (Tobin et al. 2007). 
A different scenario can be also considered, where the hot and warm \wat\,
components are actually produced in two separate non-dissociative shocks
having different velocities. Slow C-type shocks, with velocities $\la$15\kms\, 
produce post-shocked temperatures 
that never exceed $\sim$ 300-400 K
(e.g. Kaufman \& Neufeld 1996): at such temperatures, the conversion of oxygen into water 
proceeds at very low efficiency and therefore the \wat\, abundance does not
dramatically increase relative to its
to pre-shock value on the timescale of
shock evolution. In addition, as discussed before, at such low velocities 
ice mantles are not efficiently sputtered; therefore the release of water
from grains is also inhibited.
Such a scenario would, 
however, imply that the bulk of the 557 GHz line that we observe originates in a shock
having a speed much lower than the actual velocity as measured from
the line profile. 

\section{Conclusions}

 \wat\, 2$_{12}$-1$_{01}$ and 1$_{10}$-1$_{01}$ maps of the L1448 outflow
have been analysed and compared with CO(3--2), SiO(8--7) and \h\, mid-IR lines in order to
infer the origin and properties of \wat\, emission in this 
prototypical class 0 outflow.
The main results of our analysis can be  summarized as follow:

\begin{itemize}
\item On the 12\arcsec spatial scale provided by PACS, the 179\um\, line distribution appears 
patchy, with emission peaks localized in shock spots along the outflow. Strong emission
is observed towards the L1448-C source,  
which drives the main outflow 
in the region, whose spatial
extent covers the collimated and compact molecular jet observed in the \h\, S(0) and S(1)
lines. 

\item The kinematical information provided by the 557 GHz HIFI observations reveals that water
lines trace the same velocity range as the CO gas, but present a remarkably different 
profile, which is dominated by emission at intermediate velocities (i.e. $\pm$ 10-30\kms).
Emission from gas at extreme velocities (i.e. up to 80\kms) is detected but it is not
as prominent as in CO. We analyzed the velocity dependence of the \wat/CO(3--2)
ratio, finding that this ratio varies significantly with velocity. An initial 
\wat/CO(3--2) increase is followed by a drop at 
velocity $\sim$ 30\kms .
Such velocity variations are indicative of strong 
changes in the 
physical and chemical conditions with the flow speed, 
and cannot be explained  by \wat\, abundance variations alone.

\item When compared with SiO(8-7) emission, detected in our map only close to the
L1448-C source, \wat\, emission presents significant kinematical differences. SiO is
associated only with the EHV gas and it is not detected from the broad \wat\, emission 
component at intermediate velocity. The low \wat/SiO ratio inferred in the EHV bullets 
is not reproduced by shock models and points to an origin from dust-free gas directly
ejected from the protostellar wind. The absence of SiO in the broad \wat\, component 
remains 
puzzling, however, and could be explained by assuming that grain disruption
is inhibited in the very dense \wat\, emitting region.

\item From the \wat\, observed line ratio and absolute intensities, and from the
additional constraints derived from H$_2$ lines observed with \textit{Spitzer}, we infer
that the gas responsible for the bulk of the water emission is warm, 
with $T_{kin} \sim$ 300- 500 K, and very dense, with $n_{H_2}\sim$ 5\,10$^6$ \cmt.
These parameters, as well as the association of the 
179$\mu$m emission with specific shock spots, indicates that these \wat\, lines
mainly trace gas which 
has been heated and compressed 
by shocks and not entrained ambient gas, which instead mainly contributes  
to the CO(3--2) emission. 

\item The \wat\, abundance of the gas component traced by the 2$_{12}$-1$_{01}$ and 1$_{10}$-1$_{01}$
lines has been directly measured comparing the \wat\, column density with the \h\, column 
density inferred from the \h\, S(1) 17\um\, line: values of the order of 
0.5--1$\times$10$^{-6}$ are found, with small variations along the outflow, 
but these
increase by roughly an order of magnitude towards the L1448-C source. 
Such a low abundance value, associated with warm gas at a few hundred K,
 suggests that a diffuse FUV field  
may act to dissociate the freshly 
 formed water in the post-shock cooling regions. Alternative possibilities, like
 \wat\, formation in very low-velocity C-type shocks, or freeze-out of \wat\,
 molecules on dust grains in the post-shocked gas, seem 
to provide a less compelling explanation of
our findings.

\end{itemize}


\begin{acknowledgements}
The Italian authors 
gratefully acknowledge the support from ASI through the contract
I/005/011/0. Astrochemistry in Leiden is supported by NOVA, by a Spinoza grant and
grant 614.001.008 from NWO, and by EU FP7 grant 238258.  
The US authors gratefully acknowledge the support of NASA funding provided through an award
issued by JPL/Caltech.
HIFI has been designed and built by a consortium of 
institutes and university departments from across Europe, Canada and the 
United States under the leadership of SRON Netherlands Institute for Space
Research, Groningen, The Netherlands and with major contributions from 
Germany, France and the US. Consortium members are: Canada: CSA, 
U.Waterloo; France: CESR, LAB, LERMA, IRAM; Germany: KOSMA, 
MPIfR, MPS; Ireland, NUI Maynooth; Italy: ASI, IFSI-INAF, Osservatorio 
Astrofisico di Arcetri- INAF; Netherlands: SRON, TUD; Poland: CAMK, CBK; 
Spain: Observatorio Astron{\'o}mico Nacional (IGN), Centro de Astrobiolog{\'i}a 
(CSIC-INTA). Sweden: Chalmers University of Technology - MC2, RSS $\&$ 
GARD; Onsala Space Observatory; Swedish National Space Board, Stockholm 
University - Stockholm Observatory; Switzerland: ETH Zurich, FHNW; USA: 
Caltech, JPL, NHSC.

\end{acknowledgements}

\newpage

\appendix

\section{Comparison of different tracers around the L1448-C and N regions}

\subsection{L1448-C}

\begin{figure}
\centering
\includegraphics[angle=-90,width=8cm]{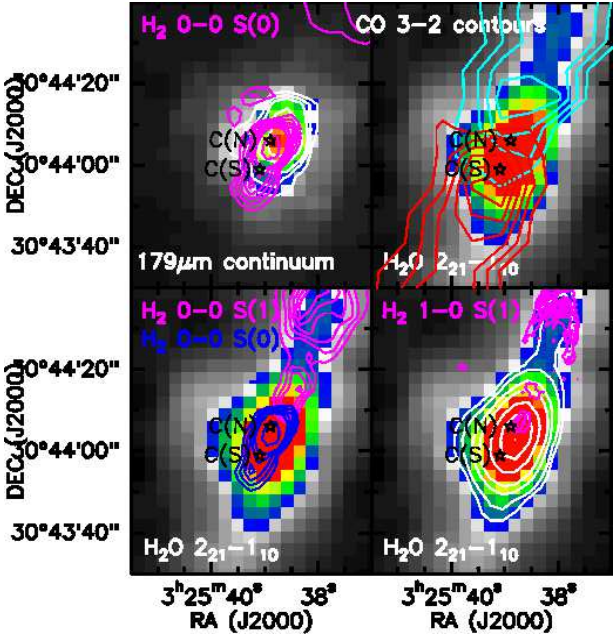}
\caption{The four panels show maps of the 179\um\, line and continuum emission 
in the region around L1448-C sources, compared with CO(3-2) and \h\, lines
at different excitation, namely the 0--0 S(0)(28\um), 0--0 S(1)(17\um), and 
1--0 S(1)(2.12\um). White contours, shown in the bottom right and upper left panels,
for the line and continuum emission respectively, are drawn at the following values:
1.5, 2.0, 3.0, 4.0, 6.0 8.0 $\times$ 10$^{-5}$ \escs\, for the line emission, and 
0.6, 0.8, 1.2, 1.6 $\times$ 10$^{-3}$ \escs\, for the continuum emission.}
\label{map_center}
\end{figure}

Fig. \ref{map_center} shows an enlargement of the 179\um\, emission in the region 
around L1448-C. Both the continuum and the \wat\, emission are displayed, with 
superimposed contours of the CO(3-2) and different \h\, lines (Near-IR, from Davis \& Smith 1996, and 
\textit{Spitzer} from Giannini et al. 2011). 
The 179\um\, line peaks towards the C(N) source but it is elongated 
along the direction of the molecular jet, as discussed in section 4.1, which in the figure is 
traced by the \h\ 0--0 S(0) 
and S(1) lines and comprises the inner peaks in the CO(3-2) emission.
The \h\, S(0) line is observed on source and along the SE (red-shifted) jet, 
while the S(1) line is detected only towards the NW blue-shifted jet and towards the B1 region.
Extinction is the likely reason why the S(1) line is not detected on-source. 
Assuming a temperature of the order of 300 K, the ratio between the S(0) flux and the S(1) 
upper limit implies a lower limit of A$_v \sim$ 65 mag towards the 
source and 45 mag in the red-shifted jet. 

Finally, Fig. \ref{map_center} shows the overlay with the \h\, 2.12\um\, line. 
At the central source position the line is almost totally extincted and thus no NIR
emission is associated with the jet. The 2.12\um\, line emission traces instead 
a bow shock in the blue lobe originated in the interaction of the jet with the ambient medium, which shows up also as a clump of \wat\, emission.  

\subsection{L1448-N}

\begin{figure}
\centering
\includegraphics[angle=-90,width=8cm]{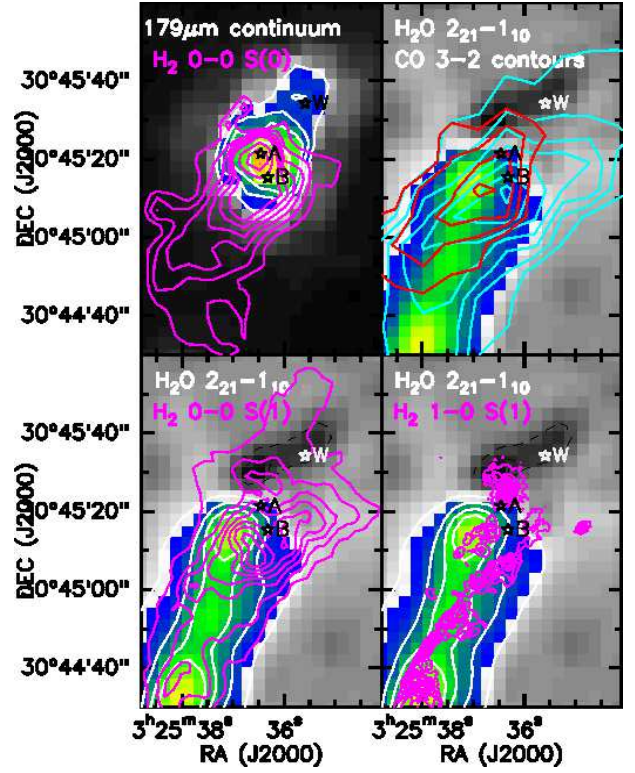}
\caption{The same as Fig. \ref{map_center} for a region around
L1448-N. In this case, white contours are drawn 
for the line and continuum emission with the following values:
0.6, 1.2, 1.8, 2.4, 4.0  $\times$ 10$^{-5}$ \escs\, (line)
and 0.6, 0.8, 1.2, 1.6 $\times$ 10$^{-3}$ \escs\, (continuum). A 
contour of black broken line delineates the absorption region.}
\label{map_north}
\end{figure}

The 179\um\, continuum image, displayed in Fig. \ref{map_north}, shows not-resolved
emission from the three sources of the L1448-N cluster, having the peak coincident with 
the N(A) source. Also the \h\, S(0) emission, 
overlaid on the continuum image,
peaks towards N(A), indicating large columns of cold gas.
As described in Sect. 3.1, only two of the three sources power outflows, resolved through interferometric observations by Girart et al. (2001) and Kwon et al. (2006). The outflow from 
N(A) is very compact and it is seen almost perpendicular to the line of sight. 
By contrast, the outflow from N(B) is more elongated and extends to about 100\arcsec\,
from the driving source (at PA=110$\degr$) both in the blue- and red-shifted
lobes. In our CO map we cannot distinguish the blue-shifted gas of these 
two outflows from the large scale L1448-C main outflow; 
however, we identify 
red-shifted emission at velocity between $\sim$ +1 and +20 \kms\, 
mainly originating from the N(B) flow (see also Fig. 3, left).

In contrast with L1448-C, the 179\um\, line emission does not peak
towards the sources of this region, but is associated only with the 
outflow: bright emission is, in particular, observed close to the H$_2$ S(1)
and to the CO(3-2) red-shifted peaks. The bulk of the water emission, 
however, does not follow the curving \h\, large scale jet driven by  
L1448-C, but seems to be associated with the 2.12\um\, \h\, emission 
(knots Y/Z in Davis \& Smith 1996), excited in the L1448-N(A/B) outflows. 
This could be a density effect, if one assumes that the density at the 
base of the N(A/B) flows, is higher than the gas along the large 
scale jet. 

North  
of the N(A/B) sources the water emission decreases abruptly, 
while an
absorption line of water appears, 
which follows the 179\um\, continuum.
The water absorption region lies along the line of sight of the L1448-N
reflection nebula, visible in the IR images at both 2.12\um\, and in the
\textit{Spitzer} IRAC maps (Davis \& Smith 1996, Tobin et al. 2007). This evidence
suggests that the cold water in the blue-shifted outflow is seen in absorption 
against the nebula, which 
therefore lies in the background. 

\section{\wat\, 557 GHz spectra and velocity channel maps}

\begin{figure*}
\centering
\includegraphics[angle=0,width=16cm]{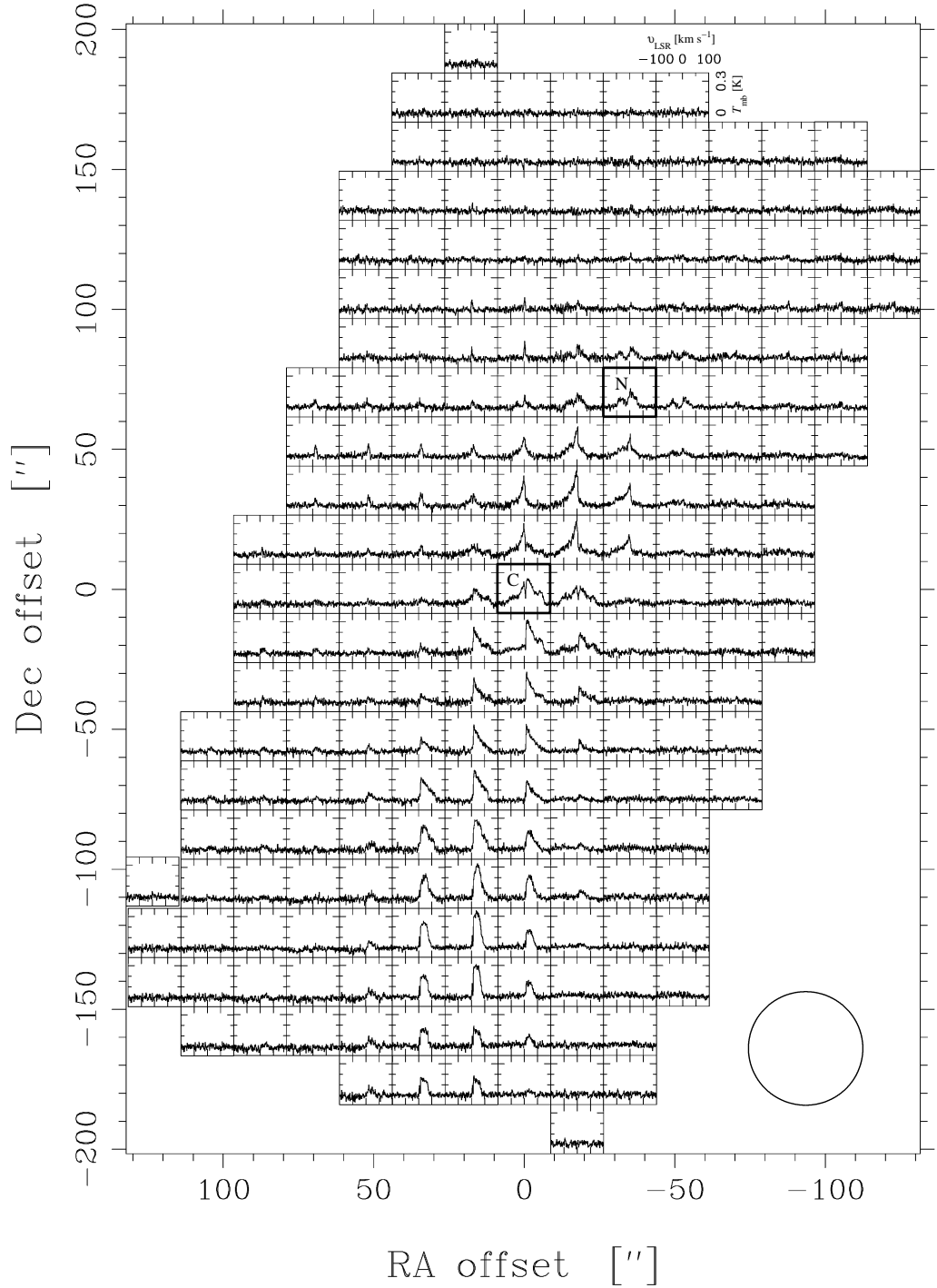}
\caption{HIFI map of the \wat\, 1$_{10}-1_{01}$ 557 GHz line. The data have been regridded
onto a regular grid of 18$\arcsec$ of spacing (i.e. half of the instrumental HPBW,
which is displayed in the figure as reference) and binned at 1\kms\, 
resolution. 
The map is centered on the L1448-C source at $\alpha_{2000}$ = 03$^h$25$^m$38.4$^s$,
$\delta_J2000$ = +30$^o$44$\arcmin$06$\arcsec$.}
\label{HIFImap}%
\end{figure*}

\begin{figure*}
\centering
\includegraphics[angle=-90,width=18cm]{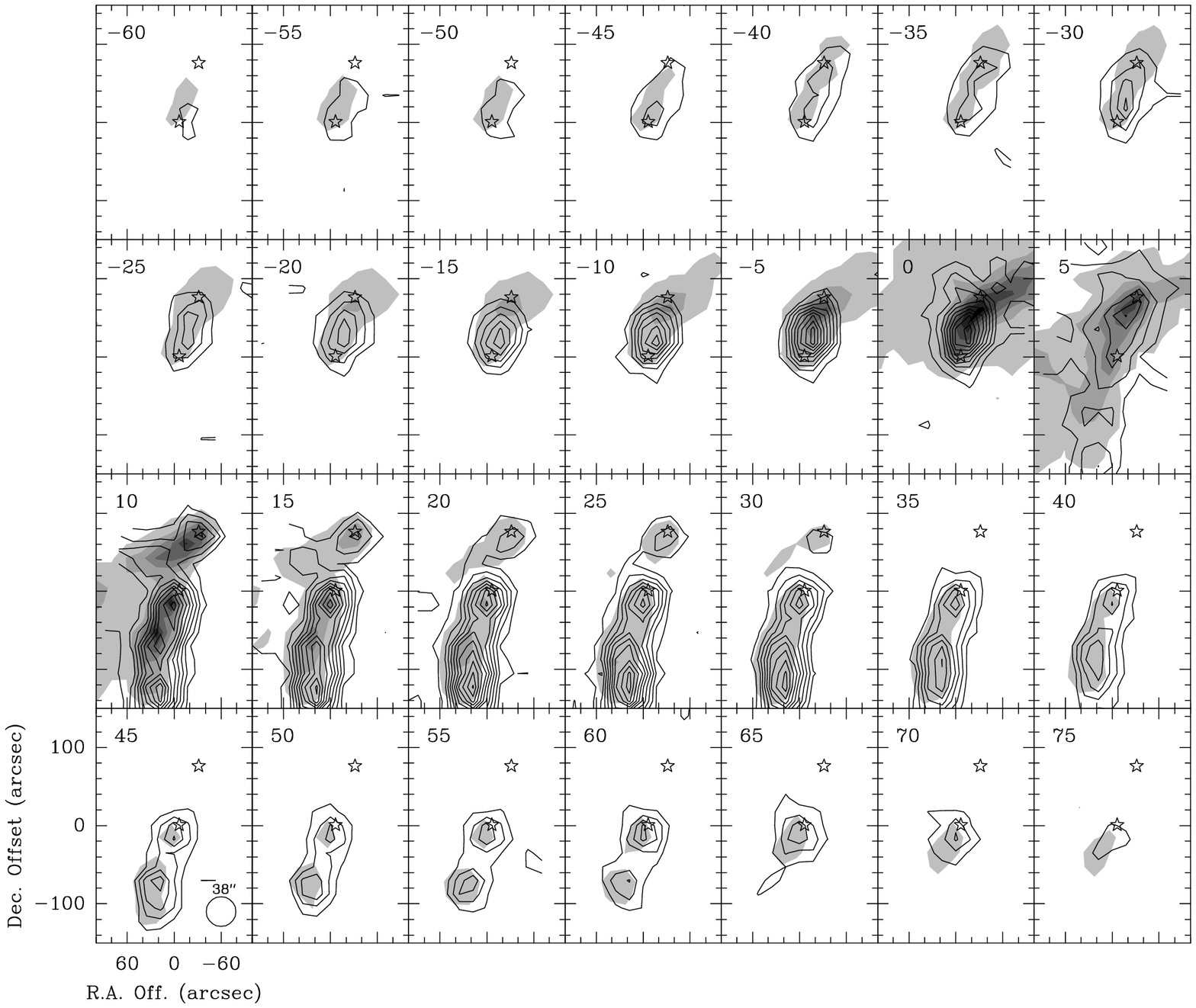}
\caption{Contours of integrated \wat 557 GHz intensities in velocity intervals of $\Delta V$ = 5 \kms\, superimposed on 
gray scale maps of the CO(3-2) intensity integrated in the same bins. 
The center velocity of the bins is indicated in each panel. \wat\, contours are in steps of 0.03 K\,\kms\,
and the first contour is at 0.03 K\kms\,(equivalent to 3$\sigma$), while CO grey levels are in steps of 1.2 K\,\kms\,
and the first contour is at 0.2  K\kms.
The starred symbols represent the positions of the L1448-C and L1448-N sources.}

\label{HIFImap}%
\end{figure*}
\end{document}